\documentclass[aps,prd,a4paper,twocolumn,longbibliography,amsmath,showpacs,superscriptaddress,nofootinbib,preprintnumbers]{revtex4-1}

\pdfoutput=1

\usepackage{amsmath}
\usepackage{amssymb}
\usepackage[american]{babel}
\usepackage{booktabs}
\usepackage{dcolumn}  
\usepackage[T1]{fontenc}  
\usepackage{graphicx}  
\usepackage[]{hyperref}  
\usepackage[utf8]{inputenc}  
\usepackage{url}
\usepackage[dvipsnames]{xcolor}
\usepackage{float}
\DeclareUnicodeCharacter{2212}{-}
\usepackage{ulem} 

\usepackage{tikz}

\usetikzlibrary{arrows,shapes,positioning,shadows,trees}

\tikzset{
	basic/.style  = {draw, text width=2cm, drop shadow, font=\sffamily, rectangle},
	root/.style   = {basic, rounded corners=2pt, thin, align=center,
		fill=green!30},
	level 2/.style = {basic, rounded corners=6pt, thin,align=center, fill=green!60,
		text width=8em},
	level 3/.style = {basic, thin, align=left, fill=pink!60, text width=6.5em}
}

\usepackage{hyperref}
\hypersetup{colorlinks=true,linkcolor=royalblue,citecolor=magenta}

\usetikzlibrary{arrows,decorations.pathmorphing,backgrounds,fit,positioning,shapes.symbols,chains}

\newcolumntype{d}[1]{D{.}{.}{#1}}
\newcolumntype{v}[1]{D{,}{,\ }{#1}}

\makeatletter

\newcommand{\Rmnum}[1]{\expandafter\@slowromancap\romannumeral #1@}

\makeatother

\renewcommand {\arraystretch}{1.3}

\hypersetup{
	citecolor = red,
	linkcolor = blue,
	urlcolor = magenta
}

\begin{document}

\title{When Dark Matter Heats Up: A Model-Independent Search for Non-Cold Behavior}

\author{Mazaharul Abedin}
\email{mazaharul.rs@presiuniv.ac.in}
\affiliation{Department of Mathematics, Presidency University, 86/1 College Street,  Kolkata 700073, India}

\author{Luis A. Escamilla}
\email{luis.escamilla@icf.unam.mx}
\affiliation{Department of Physics, Istanbul Technical University, Maslak 34469 Istanbul, T\"{u}rkiye}
\affiliation{School of Mathematical and Physical Sciences, University of Sheffield, Hounsfield Road, Sheffield S3 7RH, United Kingdom}

\author{Supriya Pan}
\email{supriya.maths@presiuniv.ac.in}
\affiliation{Department of Mathematics, Presidency University, 86/1 College Street,  Kolkata 700073, India}
\affiliation{Institute of Systems Science, Durban University of Technology, Durban 4000, Republic of South Africa}

\author{Eleonora Di Valentino}
\email{e.divalentino@sheffield.ac.uk}
\affiliation{School of Mathematical and Physical Sciences, University of Sheffield, Hounsfield Road, Sheffield S3 7RH, United Kingdom}

\author{Weiqiang Yang}
\email{d11102004@163.com}
\affiliation{Department of Physics, Liaoning Normal University, Dalian, 116029, P. R. China}

\begin{abstract}

This article questions the common assumption of cold dark matter (DM) by exploring the possibility of a non-zero equation of state (EoS) without relying on any parametric approach. In standard cosmological analyses, DM is typically modeled as pressureless dust with $w_{\rm DM} = 0$, an assumption that aligns with large-scale structure formation, supports the empirical success of the $\Lambda$CDM model, and simplifies cosmological modeling. However, there is no fundamental reason to exclude a non-zero $w_{\rm DM}$ from the cosmological framework. In this work, we explore this possibility through non-parametric and parametric reconstructions based on Gaussian Process Regression.
The reconstructions use Hubble parameter measurements from Cosmic Chronometers (CC), the Pantheon+ sample of Type Ia supernovae, and Baryon Acoustic Oscillation (BAO) data from DESI DR1 and DR2. Our findings suggest that a dynamical EoS for DM, although only mildly supported statistically, cannot be conclusively ruled out. Notably, we observe a mild tendency ($\sim 1\sigma$) toward a negative $w_{\rm DM}$ at the present epoch, which is most likely due to inconsistencies between the BAO data from DESI and other datasets.

\end{abstract}
\maketitle

\section{Introduction}\label{sec1}

In accordance with the latest observational evidence, and assuming General Relativity (GR) as the correct theory of gravity, dark matter (DM) and dark energy (DE) are the two main components of the universe, comprising nearly 96\% of its total energy budget at present moment. The $\Lambda$-Cold Dark Matter ($\Lambda$CDM) cosmological model posits that DE is most likely due to the cosmological constant ($\Lambda$), while DM is a pressureless fluid that interacts weakly with electromagnetic radiation and baryonic matter.  
This overall framework, supported by historical evidence, provides an excellent fit to a wide range of astronomical observations. However, the $\Lambda$CDM model faces several theoretical and observational challenges, and a revision of this cosmological framework may be necessary to resolve the tensions between early- and late-time cosmological observations~\cite{Verde:2019ivm,Perivolaropoulos:2021jda,Abdalla:2022yfr,DiValentino:2025sru}. Consequently, numerous modifications have been proposed in the literature to address the limitations of the standard $\Lambda$CDM model (see Refs.~\cite{DiValentino:2021izs,Schoneberg:2021qvd,Kamionkowski:2022pkx,Khalife:2023qbu,DiValentino:2020zio,Shah:2021onj,DiValentino:2022fjm,Giare:2023xoc,Hu:2023jqc,Verde:2023lmm,DiValentino:2024yew,Perivolaropoulos:2024yxv,DiValentino:2025sru} and references therein).

One possible revision of the $\Lambda$CDM cosmology, although not widely explored in the literature, is to assume a nonzero equation of state (EoS) for DM and investigate how the resulting cosmological framework may affect the expansion history of the universe and, consequently, the inferred cosmological parameters. While cosmological models with a nonzero EoS of DM (hereafter referred to as ``non-cold'' DM, following the abbreviation used in Ref.~\cite{Pan:2022qrr}) have received relatively little attention in the community~\cite{Muller:2004yb,Calabrese:2009zza,Kumar:2012gr,Armendariz-Picon:2013jej,Enqvist:2015ara,Murgia:2017lwo,Gariazzo:2017pzb,Kopp:2018zxp,Murgia:2018now,Schneider:2018xba,
Poplawski:2019tub,Ilic:2020onu,Najera:2020smt,Elizalde:2021kmo,Naidoo:2022rda,Holm:2022eqq,Khurshudyan:2023cim,
Yao:2023ybs,Khurshudyan:2024gpn,Yang:2025ume,Kumar:2025etf,Wang:2025zri,Chen:2025wwn,Li:2025eqh,Braglia:2025gdo,Kumar:2025etf}, the assumption of a nonzero DM EoS nonetheless provides a more general cosmological scenario.
It is well understood that the standard assumption of a zero EoS for DM has been motivated by the success of $\Lambda$CDM in describing large-scale structure formation and its remarkable agreement with cosmological observations. However, it is important to recognize that the intrinsic nature of both DM and DE remains unknown. As astronomical data continue to improve in precision, a more robust approach is to let observations determine the nature of DM -- whether it is truly cold or exhibits non-cold properties.
Motivated by this question, in this article we reconstruct the DM EoS using two distinct approaches: a non-parametric method and a parametric one. 
Both approaches rely on a Gaussian process framework\footnote{These approaches are ``model-independent'' in the sense that they are purely phenomenological with no assumptions on the underlying DM theoretical model.} and utilize multiple observational datasets, including Hubble parameter measurements from Cosmic Chronometers, Type Ia supernovae data, and Baryon Acoustic Oscillations.  Our results are particularly  interesting,  as they suggest  a mild indication of a dynamical evolution in the DM EoS.
While the standard CDM scenario (i.e., a DM EoS equal to zero) is recovered within the 68\% confidence level, an interesting trend emerges with the inclusion of Baryon Acoustic Oscillation data alongside other cosmological probes: a tendency toward a negative EoS at the present epoch. This suggests that the mild deviations observed in the DM EoS are best interpreted as a consequence of our phenomenological reconstructions, which act as a proxy to explore potential inconsistencies between the Dark Energy Spectroscopic Instrument (DESI) and Type Ia Supernovae datasets. In particular, since $w_{\rm DM}=0$ remains fully consistent within $2\sigma$, these results should not be overinterpreted as evidence for dynamical DM,  rather viewed as a diagnostic of the partial tension between DESI and other tracers of $\Omega_m$.

The article is organized as follows: in Section~\ref{sec-theory}, we describe the basic equations and express the EoS of DM in terms of observable quantities; in Section~\ref{sec-methodlogy}, we outline the non-parametric and parametric methodologies adopted for the reconstruction; in Section~\ref{sec-data}, we present the observational datasets used in our analysis; in Section~\ref{sec-results}, we discuss the main results and their implications; and finally, in Section~\ref{sec-summary}, we summarize the key findings of this study.

\section{Theoretical set-up}
\label{sec-theory}

On large scales, our universe is approximately homogeneous and isotropic, and this geometrical configuration is well described by the Friedmann–Lema\^{i}tre–Robertson–Walker (FLRW) metric, given by
\begin{align}
ds^2 = -dt^2 + a^2 (t) \left[ \frac{dr^2}{1- Kr^2} + r^2 \left(d\theta^2 + \sin^2 \theta\, d\phi^2 \right) \right],
\end{align}
where $K$ represents the spatial curvature of the universe and $a(t)$ is the scale factor. The curvature parameter $K$ can take the values $0, \pm 1$, corresponding to a spatially flat, open, or closed universe, respectively. In this work, we assume a spatially flat universe, i.e., $K = 0$.
In the framework of GR, the Friedmann equations governing cosmic expansion are given by
\begin{eqnarray}
   \frac{\kappa^2}{3} \rho_{\rm tot}  &=&  H^2, \label{Friedmann-1}\\
   \kappa^2 p_{\rm tot} &=&  - \left(2 \dot{H} + 3 H^2 \right),\label{Friedmann-2}
\end{eqnarray}
where $\kappa^2 = 8 \pi G$ is Einstein’s gravitational constant. The total energy density and total pressure of the cosmic fluid components, comprising DM, DE, and baryons (b), are denoted by $\rho_{\rm tot} = \rho_{\rm DM} + \rho_{\rm DE} + \rho_{\rm b}$ and $p_{\rm tot} = p_{\rm DM} + p_{\rm DE} + p_{\rm b}$, respectively. Here, $\rho_i$ and $p_i$ correspond to the energy density and pressure of the $i$-th fluid ($i = \mathrm{DM}, \mathrm{DE}, \mathrm{b}$).
Since the observational data used in this work primarily correspond to late times ($a \sim 1$), we omit the contribution of radiation to the total energy density and pressure, as it is negligible in this regime. An overhead dot denotes differentiation with respect to cosmic time, and $H \equiv \dot{a}/a$ is the Hubble rate of the FLRW universe.
We assume that each fluid follows a barotropic EoS, given by $w_i = p_i / \rho_i$. Specifically, we take $w_{\rm b} = 0$, while the DE EoS $w_{\rm DE}$ can be treated either as constant or dynamical, depending on specific parametrizations. The primary focus of this work is the DM EoS $w_{\rm DM}$, which we reconstruct using observational data.

The contracted Bianchi identity, or equivalently, the combination of equations~(\ref{Friedmann-1}) and~(\ref{Friedmann-2}), leads to the conservation equation for the total energy density:
\begin{eqnarray}
    \dot{\rho}_{\rm tot} + 3 H (\rho_{\rm tot} + p_{\rm tot}) &=& 0. \label{cons-tot}
\end{eqnarray}
Since we assume no interaction between the different fluid components, this equation can be decoupled into three separate conservation equations:
\begin{eqnarray}
    \dot{\rho}_i + 3 H (1 + w_i) \rho_i = 0, \quad (i = {\rm DM}, {\rm DE}, {\rm b}).
\end{eqnarray}
Next, we introduce the dimensionless Hubble parameter,
\begin{eqnarray}
    E(z) = \frac{H(z)}{H_0},
\end{eqnarray}
where $H_0$ is the present-day value of the Hubble parameter. Using this definition, equation~(\ref{Friedmann-1}) can be rewritten as
\begin{eqnarray}
    E^2(z) = \Omega_{\rm b,0} (1 + z)^3 + \widetilde{\Omega}_{\rm DM}(z) + \widetilde{\Omega}_{\rm DE}(z), \label{dim-relation} 
\end{eqnarray}
where $1 + z = a_0/a$ is the redshift (with $a_0$, the present-day scale factor, set to unity), and $\Omega_{\rm b,0} = 0.04$\footnote{The parameter $\Omega_{\rm b,0}$ is fixed only in one of the reconstruction approaches (which we refer to as the kernel approach), while in the other we apply a BBN Gaussian prior~\cite{Schoneberg:2024ifp} to infer the value of $\Omega_{\rm b,0}h^2$.} represents the present-day baryon density parameter. The terms $\widetilde{\Omega}_{\rm DM} = \rho_{\rm DM}/\rho_{\rm c,0}$ and $\widetilde{\Omega}_{\rm DE} = \rho_{\rm DE}/\rho_{\rm c,0}$ denote the dimensionless density parameters for DM and DE, respectively, where $\rho_{\rm c,0} = \frac{3H_0^2}{\kappa^2}$ is the present-day critical energy density.
The conservation equation for DE, given by
\begin{eqnarray}    
\dot{\rho}_{\rm DE} + 3 H (1+w_{\rm DE}) \rho_{\rm DE} = 0,
\end{eqnarray}
can be rewritten in terms of the redshift as
\begin{eqnarray}    
\frac{3(1+w_{\rm DE})}{(1+z)}\widetilde{\Omega}_{\rm DE} = \widetilde{\Omega}_{\rm DE}', \label{Omega-tilde-de}
\end{eqnarray}
where the prime denotes differentiation with respect to redshift $z$.
Next, taking the derivative of equation~(\ref{dim-relation}) with respect to $z$ and using equation~(\ref{Omega-tilde-de}), we obtain the evolution equation for the DM density:
\begin{align}
 \widetilde{\Omega}_{\rm DM}' = 2E E' - 3 \Omega_{\rm b,0} (1+z)^2 - \frac{3(1+w_{\rm DE})}{(1+z)} \widetilde{\Omega}_{\rm DE}. \label{derivative-Omega-tilde-DM} 
\end{align}
Finally, using the conservation equation for DM,
\begin{eqnarray}
\dot{\rho}_{\rm DM} + 3 H (1+ w_{\rm DM}) \rho_{\rm DM} = 0,
\end{eqnarray}
we derive the EoS for DM, $w_{\rm DM}$, as
\begin{eqnarray}
    w_{\rm DM} = -1 + \frac{(1+z) \widetilde{\Omega}_{\rm DM}'}{3 \widetilde{\Omega}_{\rm DM}}, \label{EoS-DM}
\end{eqnarray}
which explicitly shows that the evolution of $w_{\rm DM}$ depends on the redshift evolution of the other cosmic fluids.
In this article, we aim to investigate the key question: 
 whether the evolution of $w_{\rm DM}$ exhibits any deviation from $w_{\rm DM} = 0$ from a non-parametric perspective.  
Using the previously derived equations, we can express $w_{\rm DM}$ as follows:
\begin{eqnarray}\label{eos-DM-rec-E}
   w_{\rm DM} &=& \frac{2E E'(1+z) - 3E^2 - 3w_{\rm DE} \widetilde{\Omega}_{\rm DE}}{3\left[E^2 - \Omega_{\rm b,0}(1+z)^3 - \widetilde{\Omega}_{\rm DE} \right]}. 
\end{eqnarray}

In terms of the dimensionless comoving distance, $D(z) = \int_0^z \frac{dz'}{E(z')}$, the EoS of DM can be written as
\begin{eqnarray}\label{eos-DM-rec-D}
   w_{\rm DM} &=& \frac{-2D''(1+z) - 3D' - 3w_{\rm DE} D'^3 \widetilde{\Omega}_{\rm DE}}{3\left[D' - D'^3 \Omega_{\rm b,0}(1+z)^3 - D'^3 \widetilde{\Omega}_{\rm DE} \right]}. 
\end{eqnarray}

Now, regarding the DE sector that appears in equations~(\ref{eos-DM-rec-E}) and~(\ref{eos-DM-rec-D}) through the term $\widetilde{\Omega}_{\rm DE}$, one can consider either a constant or a dynamical EoS for DE. However, in this work, we adopt the simplest case by setting $w_{\rm DE} = -1$, corresponding to a vacuum energy density, and thus $\widetilde{\Omega}_{\rm DE}$ becomes constant and actually represents the present-day value of the DE density parameter, i.e., $\widetilde{\Omega}_{\rm DE}(z = 0) = \Omega_{\rm DE,0}$.

\section{Methodology}
\label{sec-methodlogy}

\subsection{Gaussian process with kernels}
\label{sec-Gaussian-approach}

The relationship (or regression) between $z$ and $w_{\rm DM}(z)$ can be determined using a Gaussian Process (GP), a non-parametric approach~\cite{2012JCAP...06..036S,books/lib/RasmussenW06,Mukherjee:2021ggf,Velazquez:2024aya}, without assuming any specific parametric form.
Using Gaussian Process Regression (GPR), we reconstruct the functions $E(z)$ and $D(z)$, along with their corresponding derivatives, directly from the data. These reconstructed quantities are then employed in Eqs.~\eqref{eos-DM-rec-E} and \eqref{eos-DM-rec-D} to compute the EoS of DM, $w_{\rm DM}(z)$.
The uncertainty in $w_{\rm DM}(z)$ is estimated through an error propagation technique that accounts for the statistical properties of $E(z)$ and $D(z)$, which follow a multivariate Gaussian distribution. These properties are characterized by their mean functions and covariance matrices, including the observational noise associated with $E(z)$ and $D(z)$.
We therefore employ various forms of covariance functions~\cite{books/lib/RasmussenW06,e6e4c2b74de0412ab4feae906164e191}. While we primarily focus on the squared exponential covariance function, we also consider alternative kernels. The squared exponential kernel, also known as the \textit{Gaussian kernel}, is given by
\begin{equation}
K(z, \tilde{z}) = \sigma_f^2 \exp\left[-\frac{(z - \tilde{z})^2}{2l^2}\right],
\label{eqn_squared_Exp}
\end{equation}
where $z$ and $\tilde{z}$ are two points, and $\sigma_f$ and $l$ are hyperparameters controlling the vertical variation and the characteristic length scale, respectively. This kernel is widely used due to its smoothness and the fact that it is infinitely differentiable.
Another commonly used kernel is the Matérn kernel, defined as
\begin{align}
\label{cov_Matern}
K_{\nu = p + \frac{1}{2}}(z,\tilde{z}) &= \sigma_f^2 \exp\left[-\frac{\sqrt{2p + 1}}{l} \vert z - \tilde{z} \vert \right] \frac{p!}{(2p)!} \nonumber\\
&\quad \times \sum_{i=0}^{p} \frac{(p + i)!}{i!(p - i)!} \left( \frac{2\sqrt{2p + 1}}{l} \vert z - \tilde{z} \vert \right)^{p - i}.
\end{align}
For each value of $p$, a unique Matérn covariance function is obtained. Examples include Matérn $9/2$ for $p = 4$, Matérn $7/2$ for $p = 3$, Matérn $5/2$ for $p = 2$, and Matérn $3/2$ for $p = 1$. These Matérn covariance functions are $r$-times mean-square differentiable if $r < \nu$, where $\nu = p + \frac{1}{2}$.
Additionally, we consider another kernel, known as the Cauchy kernel, defined as
\begin{equation}\label{Cauchy_cov}
K(z, \tilde{z}) = \sigma_f^2 \left(\frac{l}{(z - \tilde{z})^2 + l^2}\right),
\end{equation}
and the Rational Quadratic kernel, given by
\begin{equation}\label{Rational_cov}
K(z, \tilde{z}) = \sigma_f^2 \left(1 + \frac{(z - \tilde{z})^2}{2\alpha l^2}\right)^{-\alpha},
\end{equation}
where $\sigma_f$, $\alpha$, and $l$ are hyperparameters.

In Appendix~\ref{sec-app-A}, we describe the methodology of the Gaussian Process in detail. 
To perform GPR, we use the Python package \href{https://github.com/astrobengaly/GaPP}{\texttt{GaPP}},\footnote{\url{https://github.com/astrobengaly/GaPP}} which also allows the computation of the second and third derivatives of the GPR reconstruction. 
Additionally, we employ a modified version of \texttt{GaPP} to handle combined datasets. The methodology is discussed in~\cite{Dinda:2024ktd}.

\subsection{Gaussian Process as an Interpolation}

We also adopt a parametric approach to reconstruct $w_{\rm{DM}} (z)$. This method employs a Gaussian process, though not in a fully non-parametric manner, as it involves parameters whose values must be inferred from data using Bayesian statistics. In particular, we use a parameter inference method known as Nested Sampling~\cite{skilling}. In this way, we can directly compare our parametric reconstruction with the standard model using the Bayesian evidence and the maximum log-likelihood. For a brief review of Bayesian statistics, see Appendix~\ref{sec-app-B}. 

In this manner, our parametric reconstruction consists of a set of ``nodes,'' which are interpolated using a Gaussian process. These nodes vary in height (their ordinate values), and these variable heights serve as the parameters of the reconstruction. For $n$ nodes, there are $n$ variable heights, resulting in $n$ additional parameters.
This method has previously been applied to reconstruct the EoS of DE~\cite{Gerardi:2019obr} and 
the interaction rate of an interacting DE model  
~\cite{Escamilla:2023shf}.

The sampler code used in this work is a modified version of an MCMC code named \texttt{SimpleMC}~\cite{simplemc}. \texttt{SimpleMC} is designed to calculate expansion rates and distances based on the Friedmann equation. It also employs a Nested Sampling library, \texttt{dynesty}~\cite{2020MNRAS.493.3132S}, to compute the Bayesian evidence during the parameter inference procedure.
Regarding the priors on the parameters, we adopt the following ranges: $\Omega_m = [0.1, 0.5]$ for the matter density parameter, $\Omega_b h^2 = [0.02, 0.025]$ for the physical baryon density, $h = [0.4, 0.9]$ for the dimensionless Hubble parameter, where $H_0 = 100h$ km s$^{-1}$ Mpc$^{-1}$, and $w_{\rm DM,i} = [-5.0, 5.0]$ for the nodes that define the DM parametric EoS. In this work, the number of nodes is set to 5.
The number of live points used in the Nested Sampling algorithm is chosen based on the general rule $50 \times ndim$~\cite{dynestyy}, where $ndim$ is the number of parameters being sampled.

\begin{table}[!]
\begin{center}
\resizebox{0.47\textwidth}{!}{%
    \tiny 
\renewcommand{\arraystretch}{1.0} 
\begin{tabular}{|l c c r|l c c r|}
\multicolumn{8}{c}{}\\
\hline
$z$ & $H(z)$  & $\sigma_{H(z)}$  & Ref. ~& $~z$ & $H(z)$ & $\sigma_{H(z)}$  & Ref.\\
\hline

0.07 &   69.0   &  19.6 &  \cite{Zhang:2012mp}~ &~0.4783 &   80.9   &  9.0 &  \cite{Moresco:2016mzx}\\ 
0.09 &   69.0   &  12.0 &  \cite{Jimenez:2003iv}~ & ~0.48 &   97.0   &  62.0 &  \cite{Stern:2009ep} \\
0.12 &   68.6   &  26.2 &  \cite{Zhang:2012mp}~ & ~0.5929 &   104.0   &  13.0 &  \cite{Moresco:2012by} \\ 
0.17 &   83.0   &  8.0 &  \cite{Simon:2004tf}~ & ~0.6797 &   92.0   &  8.0 &  \cite{Moresco:2012by} \\ 
0.1791 &   75.0   &  4.0 &  \cite{Moresco:2012by}~ &~0.7812 &   105.0   &  12.0 &  \cite{Moresco:2012by} \\ 
0.1993 &   75.0   &  5.0 &  \cite{Moresco:2012by}~ &~0.8754 &   125.0   &  17.0 &  \cite{Moresco:2012by} \\ 
0.20 &   72.9   &  29.6 &  \cite{Zhang:2012mp}~ & ~0.88 &   90.0   &  40.0 &  \cite{Stern:2009ep} \\ 
0.27 &   77.0   &  14.0 &  \cite{Simon:2004tf}~ & ~0.90 &   117.0   &  23.0 &  \cite{Simon:2004tf} \\
0.28 &   88.8   &  36.6 &  \cite{Zhang:2012mp}~ & ~1.037 &   154.0   &  20.0 &  \cite{Moresco:2012by} \\
0.3519 &   83.0   &  14.0 &  \cite{Moresco:2012by}~ &~1.3 &   168.0   &  17.0 &  \cite{Simon:2004tf} \\
0.3802 &   83.0   &  13.5 &  \cite{Moresco:2016mzx}~ & ~1.363 &   160.0   &  33.6 &  \cite{Moresco:2015cya} \\
0.4 &   95.0   &  17.0 &  \cite{Simon:2004tf}~  & ~1.43 &   177.0   &  18.0 &  \cite{Simon:2004tf} \\ 
0.4004 &   77.0   &  10.2 &  \cite{Moresco:2016mzx}~ & ~1.53 &   140.0   &  14.0 &  \cite{Simon:2004tf} \\ 
0.4247 &   87.1   &  11.2 & ~\cite{Moresco:2016mzx}~ & ~1.75 &   202.0   &  40.0 &  \cite{Simon:2004tf} \\ 
0.4497 &   92.8   &  12.9 &  ~\cite{Moresco:2016mzx}~ &~1.965 &   186.5   &  50.4 &  \cite{Moresco:2015cya} \\ 
0.47 &   89.0   &  49.6 &  ~\cite{Ratsimbazafy:2017vga}~ &~$-$ &   $-$    & $-$ &  $-$ \\ 
\hline
\end{tabular}
}

\end{center}
\caption{We consider 31 $H(z)$ measurements obtained using the CC  method. Taking into account both statistical and systematic uncertainties, the total error is computed as $\sigma_{\rm tot} = \sqrt{\left[(\sigma_{\rm stat}^{+} + \sigma_{\rm stat}^{-})/2\right]^2 + \left[(\sigma_{\rm syst}^{+} + \sigma_{\rm syst}^{-})/2\right]^2}$. For example, the $H(z)$ value reported in~\cite{Ratsimbazafy:2017vga} is $89 \pm 23~(\mathrm{stat}) \pm 44~(\mathrm{syst})~\mathrm{km\ s^{-1}\ Mpc^{-1}}$, which corresponds to a total uncertainty of $89 \pm 49.6~(\mathrm{tot})~\mathrm{km\ s^{-1}\ Mpc^{-1}}$ in our analysis. Similarly, the measurement $113.1 \pm 15.1~(\mathrm{stat})^{+29.1}_{-11.3}~(\mathrm{syst})~\mathrm{km\ s^{-1}\ Mpc^{-1}}$ from~\cite{Jiao:2022aep} yields a total uncertainty of $113.1 \pm 25.22~(\mathrm{tot})~\mathrm{km\ s^{-1}\ Mpc^{-1}}$.  }
    \label{tab:Hz}

\end{table}

\section{Observational data}
\label{sec-data}

In this section, we describe the observational data and statistical methodology considered in this work. We outline below the datasets used for the reconstructions.

\begin{itemize}

\item {\bf Hubble parameter measurements from Cosmic Chronometers:} The Hubble parameter $H(z)$ serves as a key indicator of the properties of dark energy and dark matter, as well as the universe's expansion history. It can be estimated using a nearly model-independent method based on the differential age of passively evolving galaxies, by measuring the ages of old stellar populations at different redshifts $z$ with respect to cosmic time $t$, following the relation:
\begin{equation}
    H(z) \approx - \frac{1}{1+z} \frac{\Delta z}{\Delta t}.
\end{equation}
These old galaxies, known as Cosmic Chronometers (CC), act as ``standard clocks'' in cosmology. In this work, we use a catalogue of 31 CC measurements, listed in Table~\ref{tab:Hz}, along with their corresponding references.

\item {\bf Type Ia Supernovae:} 
The Pantheon+ sample~\cite{Brout:2022vxf} includes 1701 light curves from 1550 distinct Type Ia supernovae (SN Ia) in the redshift range $0.01 \leq z \leq 2.26$. 
To mitigate the effects of anomalous velocity corrections, we use 1590 SN Ia data points within the range $0.01 \leq z \leq 2.26$, applying the constraint $z > 0.01$~\cite{Park:2024jns}. The full Pantheon+ catalogue and its covariance matrix are publicly available.\footnote{\url{https://github.com/PantheonPlusSH0ES/DataRelease}}

The relationship between the distance modulus $\mu$ and the luminosity distance $d_{\rm L}(z)$ is given by:
\begin{equation}
    \mu = 5\log_{10} \frac{d_{\rm L}(z)}{\rm Mpc} + 25, \label{SN_dl}
\end{equation}
where $\mu = m - M_{\rm B}$ and $M_{\rm B} = -19.42$.
The comoving distance is defined as:
\begin{equation} \label{ch1:D_from_DL}
D_{\rm M}(z) \equiv \frac{d_{\rm L}(z)}{1+z},
\end{equation}
where $d_{\rm L} = 10^{\frac{(\mu - 25)}{5}}$. The error propagation rule is used to compute the covariance matrix of $D_{\rm M}(z)$.

\item {\bf Baryon Acoustic Oscillations:} Like CC and SN Ia, Baryon Acoustic Oscillations (BAO) act as ``standard rulers'' in cosmology, enabling measurements of the angular diameter distance $D_A(z)$. From this, we compute the comoving angular diameter distance $D_{\rm M} = D_A(1+z)$ and then derive $D_{\rm M}/r_{\rm d}$ using the relation:
\begin{equation}
    \frac{D_{\rm M}}{r_{\rm d}} = \frac{c}{r_{\rm d}} \int_0^z \frac{d\tilde{z}}{H(\tilde{z})}, \label{D_M_eqn}
\end{equation}
where $H(z)$ is the Hubble parameter, $c$ is the speed of light and $r_{\rm d}$ is the comoving sound horizon at the end of the baryon-drag epoch.
Since the right-hand side of Eq.~\eqref{D_M_eqn} defines the comoving distance, $D_{\rm M}(z)$, we can equivalently express $D_{\rm M}/r_{\rm d}$ as the comoving distance scaled by the sound horizon.
In this work, we use measurements of $D_{\rm M}/r_{\rm d}$ at five different redshifts. The data are taken from the Dark Energy Spectroscopic Instrument (DESI) Data Release 1 ~(DESI DR1)~\cite{DESI:2024mwx}  and Data Release 2 ~(DESI DR2)~\cite{DESI:2025zgx}  anisotropic BAO observations, with the corresponding values listed in Table~\ref{tab:DR1data} (DESI DR1) and Table~\ref{tab:DR2data} (DESI DR2).
\end{itemize}

In this work, we used a new technique to combine different datasets that measure distinct cosmological observables, which are related to each other through simple derivatives. This new technique is not exactly the same as ones used in previous studies, as it includes some modifications~\cite{Yang:2025kgc}. 
For the Pantheon+ data, we used only $z$, $D_{\rm M}(z)$, and the covariance matrix $\textit{Cov}_{D_{\rm M}}$ to reconstruct $D_{\rm M}(z)$ and its derivatives. Similarly, for CC data alone, we reconstructed $H(z)$ and its derivatives. Since DESI provides the measurements of $D_{\rm M}/r_{\rm d}$, 
we need to multiply  the DESI measurements by a specific value of $r_{\rm d}$\footnote{Note that the choice of $r_{\rm d}$ is not unique; in principle, one can adopt any particular value of $r_{\rm d}$. Since we restrict our analysis to the low-redshift regime, the methodologies applied in this work do not allow us to determine the value of $r_{\rm d}$.} to convert them into the measurements of $D_{\rm M}(z)$. 
As a result, DESI and Pantheon+ now share the same observable, namely $D_{\rm M}(z)$, and can be combined directly. However, when combining CC with DESI or Pantheon+ (or both for that matter) we must take into account  the fact that DESI and Pantheon+ provide $D_{\rm M}(z)$, while CC provide $H(z)$ measurements. For CC, the $H(z)$ data are transformed into $D_{\rm H}(z) = c/H(z)$. Since $D_{\rm H}(z)$ is the derivative of $D_{\rm M}(z)$, we can then use both types of measurements simultaneously in the GPR framework to reconstruct $D_{\rm M}(z)$ and its derivatives.
It is well known that different types of covariance functions can be used in GPR. In particular, the authors of~\cite{Seikel:2013fda,Hwang:2022hla} explored their impact on cosmological reconstructions. Following this motivation, in the present work we reconstruct $w_{\rm DM}(z)$ using both the approaches described in section~\ref{sec-Gaussian-approach}.

\begin{table}
\scalebox{0.9}{ 
    \begin{tabular}{cccccccc} 
    \hline
   Tracer & redshift  & $z_{\rm eff}$ & $D_{\rm M}/r_{\rm d}$   
     \\
    \hline
     LRG1  & $0.4-0.6$ &~ $0.510$ ~&~ $13.62\pm 0.25$~\\
    LRG2  & $0.6-0.8$ &~  $0.706$ & $16.85 \pm 0.32$ \\
    LRG3+ELG1  & $0.8-1.1$ &~ 
    $0.930$  & $21.71 \pm 0.28$   \\
    ELG2  & $1.1-1.6$ &~  $1.317$  & $27.79 \pm 0.69$  \\
    Lya QSO  & $1.77-4.16$ &~  $2.330$  & $39.71 \pm 0.94$ \\
    \hline
    \end{tabular}}
\caption{\label{tab:DR1data} 
Measurements of $D_{\rm M}/r_{\rm d}$ from DESI DR1 BAO data~\cite{DESI:2024mwx} at five distinct redshifts.}
\end{table}
\begin{table}
\scalebox{0.9}{ 
    \begin{tabular}{ccccccccc} 
    \hline
   Tracer & redshift  & $z_{\rm eff}$ & $D_{\rm M}/r_{\rm d}$    
     \\
    \hline
     LRG1 & $0.4-0.6$ &~ $0.510$ ~&~ $13.588 \pm 0.167$\\
    LRG2 &$0.6-0.8$ &~  $0.706$ &~ $17.351 \pm 0.177$ \\
    LRG3+ELG1 &$0.8-1.1$   &~ 
    $0.934$  &~ $21.576 \pm 0.152$  \\
    ELG2 & $1.1-1.6$   &~  $1.321$  &~ $27.601 \pm 0.318$  \\
    QSO & $1.1-1.6$   &~  $1.484$  &~ $30.512 \pm 0.760$ \\
    Lya & $1.77-4.16$   &~  $2.330$  &~ $38.988 \pm 0.531$ \\
    \hline
    \end{tabular}}
\caption{\label{tab:DR2data} 
Measurements of $D_{\rm M}/r_{\rm d}$ from DESI DR2 BAO data~\cite{DESI:2025zgx} at six distinct redshifts.}
\end{table}

\begin{table}[h]
    \centering
\caption{Reconstructed value of $H(z)$ at $z = 0$, i.e., $H_0$, at 68\% CL using GPR for different observational datasets.}
    \label{tab:rec_H0}
    \renewcommand{\arraystretch}{1.2} 
    \begin{tabular}{|c|c|c|}
        \hline
        \hline
        \textbf{Dataset} & \boldmath{$r_{\rm d}$(Mpc)} & \boldmath{$H_0$ (km/s/Mpc)} \\
        \hline
        CC & $-$ & $67.4\pm4.8$ \\
        Pantheon+ & $-$ & $68.74\pm0.45$ \\
        CC+Pantheon+ & $-$ & $68.78\pm0.41$ \\
        \hline
        CC+DESI-DR1 & $149.3\pm2.7$ & $71.2\pm3.5$ \\
        CC+DESI-DR1 & $137\pm3.6$ & $70.8\pm3.5$ \\
        \hline
        CC+DESI-DR2 & $149.3\pm2.7$ & $71.1\pm3.4$ \\
        CC+DESI-DR2 & $137\pm3.6$ & $70.8\pm3.6$ \\
        \hline
        DESI-DR1+Pantheon+ & $149.3\pm2.7$ & $68.72\pm0.42$ \\
        DESI-DR1+Pantheon+ & $137\pm3.6$ & $68.98\pm0.50$ \\
        \hline
        DESI-DR2+Pantheon+ & $149.3\pm2.7$ & $68.75\pm0.42$ \\
        DESI-DR2+Pantheon+ & $137\pm3.6$ & $68.90\pm0.48$ \\
        \hline
        CC+DESI-DR1+Pantheon+ & $149.3\pm2.7$ & $68.82\pm0.40$ \\
        CC+DESI-DR1+Pantheon+ & $137\pm3.6$ & $68.77\pm0.44$ \\
        \hline
        CC+DESI-DR2+Pantheon+ & $149.3\pm2.7$ & $68.823\pm0.39$ \\
        CC+DESI-DR2+Pantheon+ & $137\pm3.6$ & $68.86\pm0.45$ \\
        \hline
        \hline
    \end{tabular}
\end{table}

\section{Results}
\label{sec-results}

In this section, we present the reconstructions of $w_{\rm DM}(z)$ using various combinations of the aforementioned datasets and GPR approaches: kernel-based (non-parametric) and parametric. 
First, we reconstruct $w_{\rm DM}(z)$ using CC, Pantheon+, and CC+Pantheon+. Then, we incorporate BAO data from DESI-DR1 and DESI-DR2  and performed 
additional reconstructions with CC+DESI-DR1, DESI-DR1+Pantheon+ (DESI-DR2+Pantheon+), and CC+DESI-DR1+Pantheon+ (CC+DESI-DR2+Pantheon+),\footnote{For the BAO data from DESI, we have considered both DR1 and DR2 measurements in order to examine whether the latter makes any changes in the reconstructions compared to the former.} resulting in a total of nine dataset combinations. 
As BAO data incorporates $r_{\rm d}$, therefore, reconstructions using BAO are not fully non-parametric.
This, however, does not discourage us from examining the effects resulting from the inclusion of DESI BAO when combined with CC and Pantheon+. 
 
\begin{figure}
\includegraphics[width=0.5\textwidth]{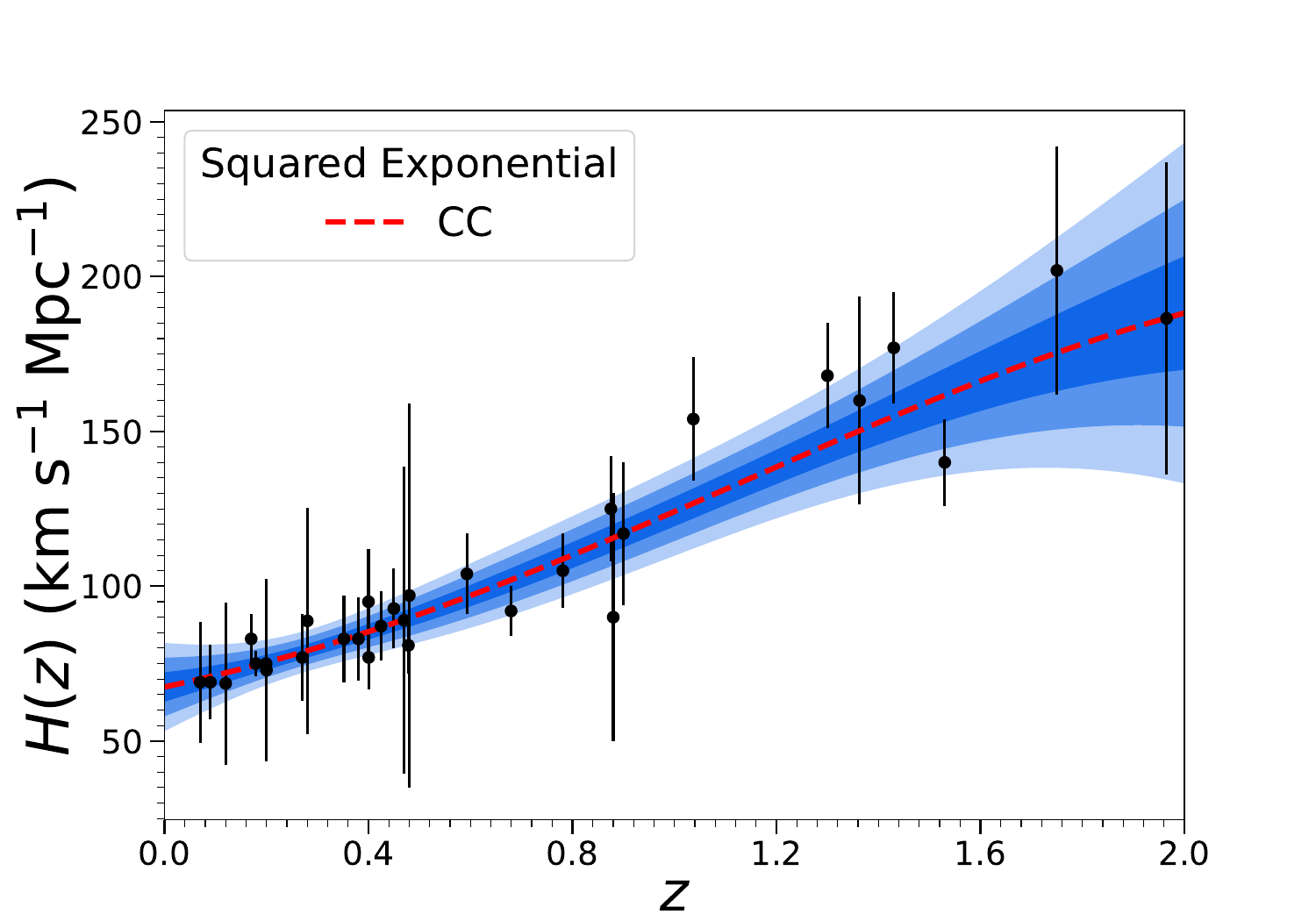}
\includegraphics[width=0.5\textwidth]{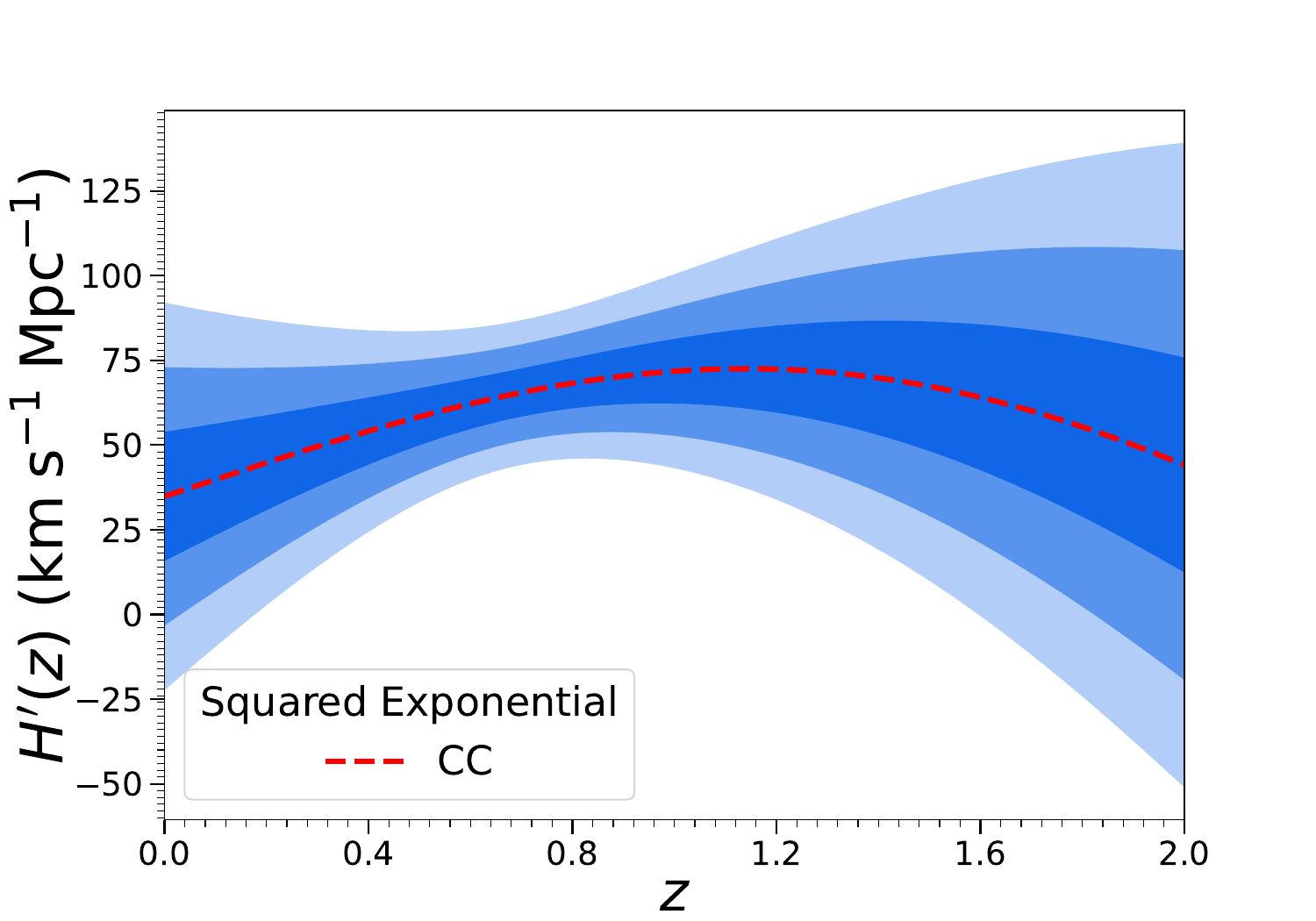}
\caption{Reconstruction of $H(z)$ (upper panel) and $H'(z)$ (lower panel) using 31 CC data points with the {\it squared exponential kernel} of the Gaussian approach. The dotted curve in each panel represents the mean of the corresponding reconstructed quantity.}
\label{fig:Hz-dHz}
\end{figure}
\begin{figure*}[t!]
    \centering
    \makebox[10cm][c]{
     \includegraphics[trim = 0mm  0mm 20mm 0mm, clip, width=5.6cm, height=4.5cm]{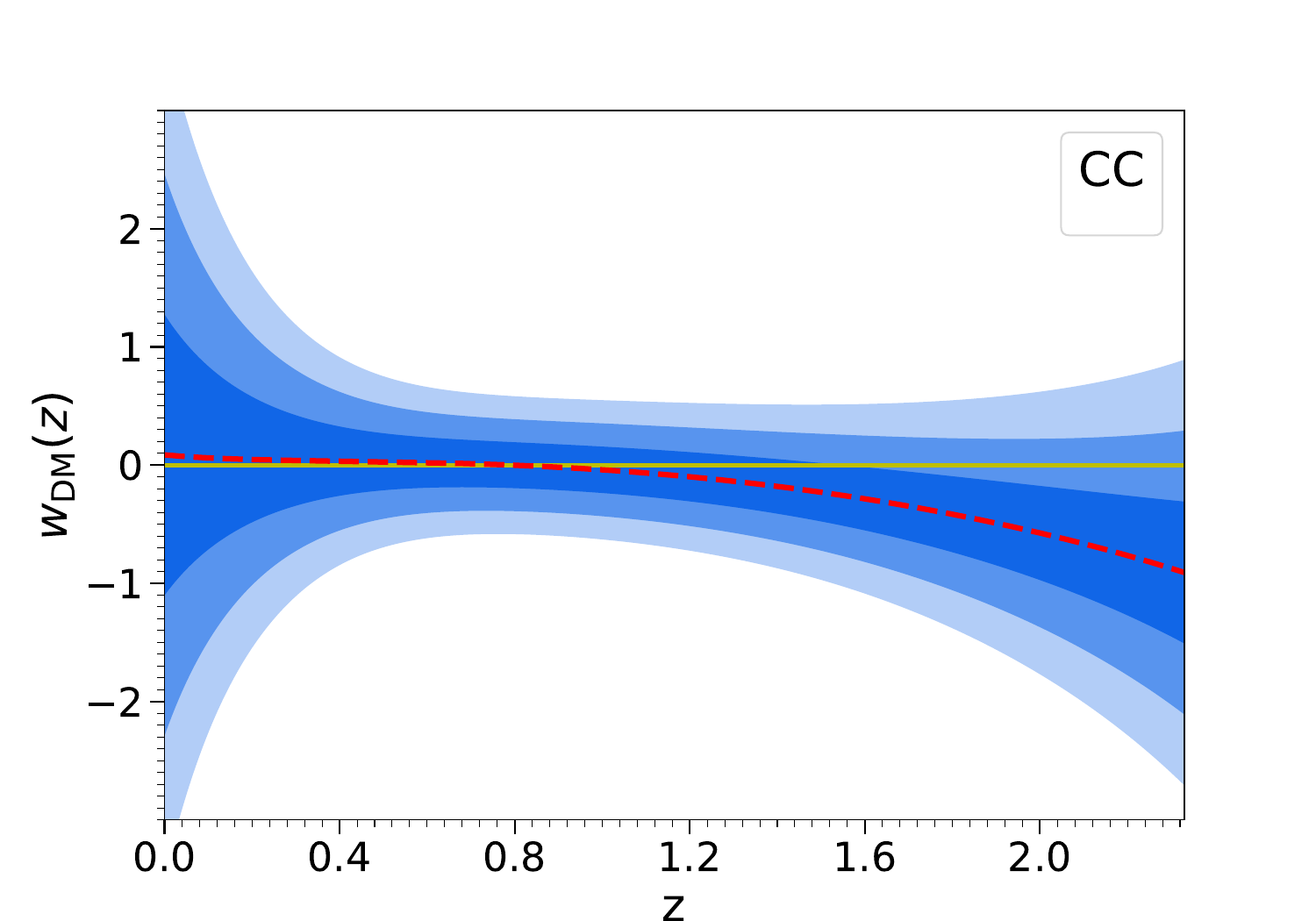}
     \includegraphics[trim = 0mm  0mm 0mm 0mm, clip, width=5.6cm, height=4.cm]{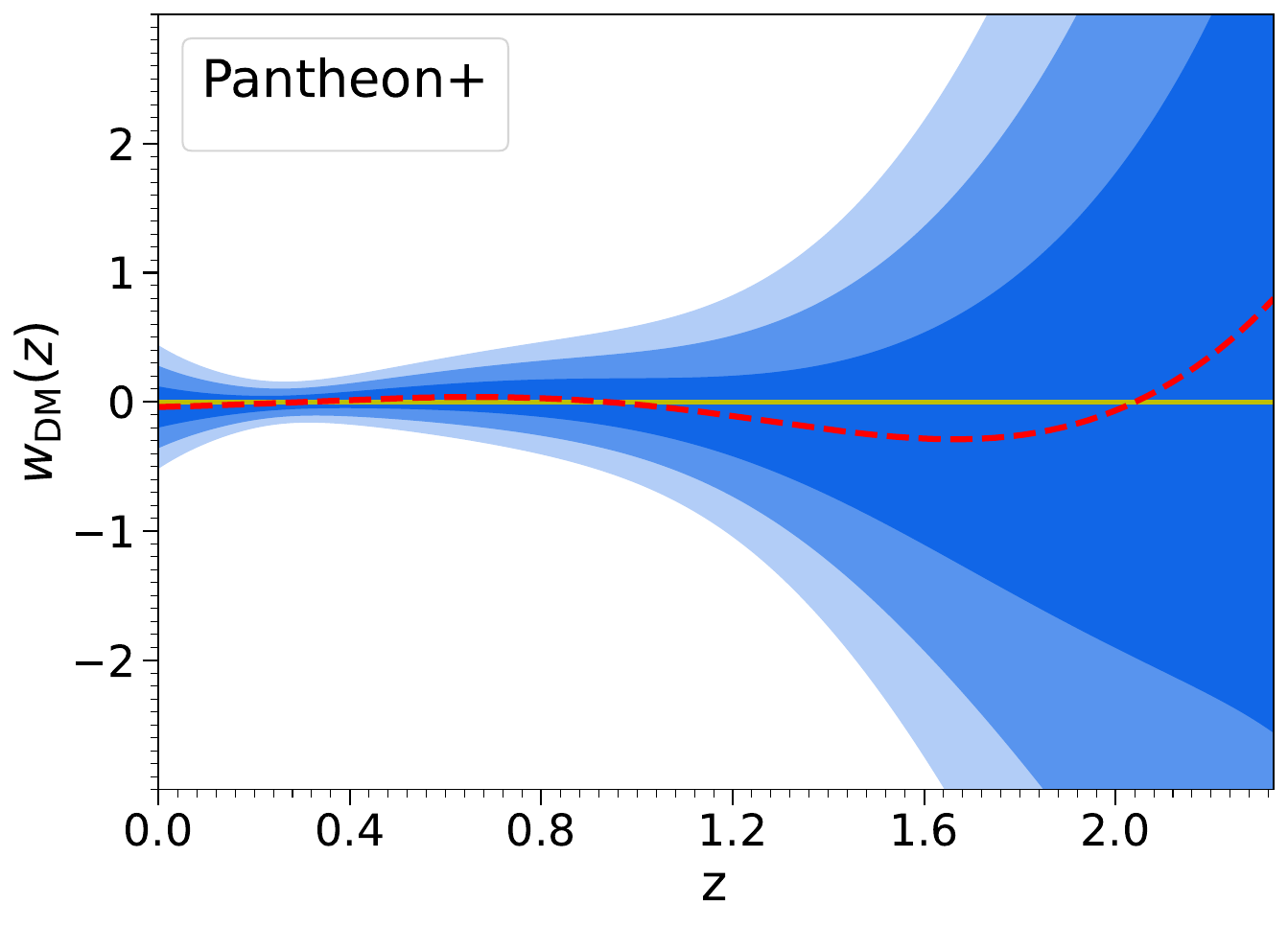}
     \includegraphics[trim = 0mm  0mm 0mm 0mm, clip, width=5.6cm, height=4.cm]{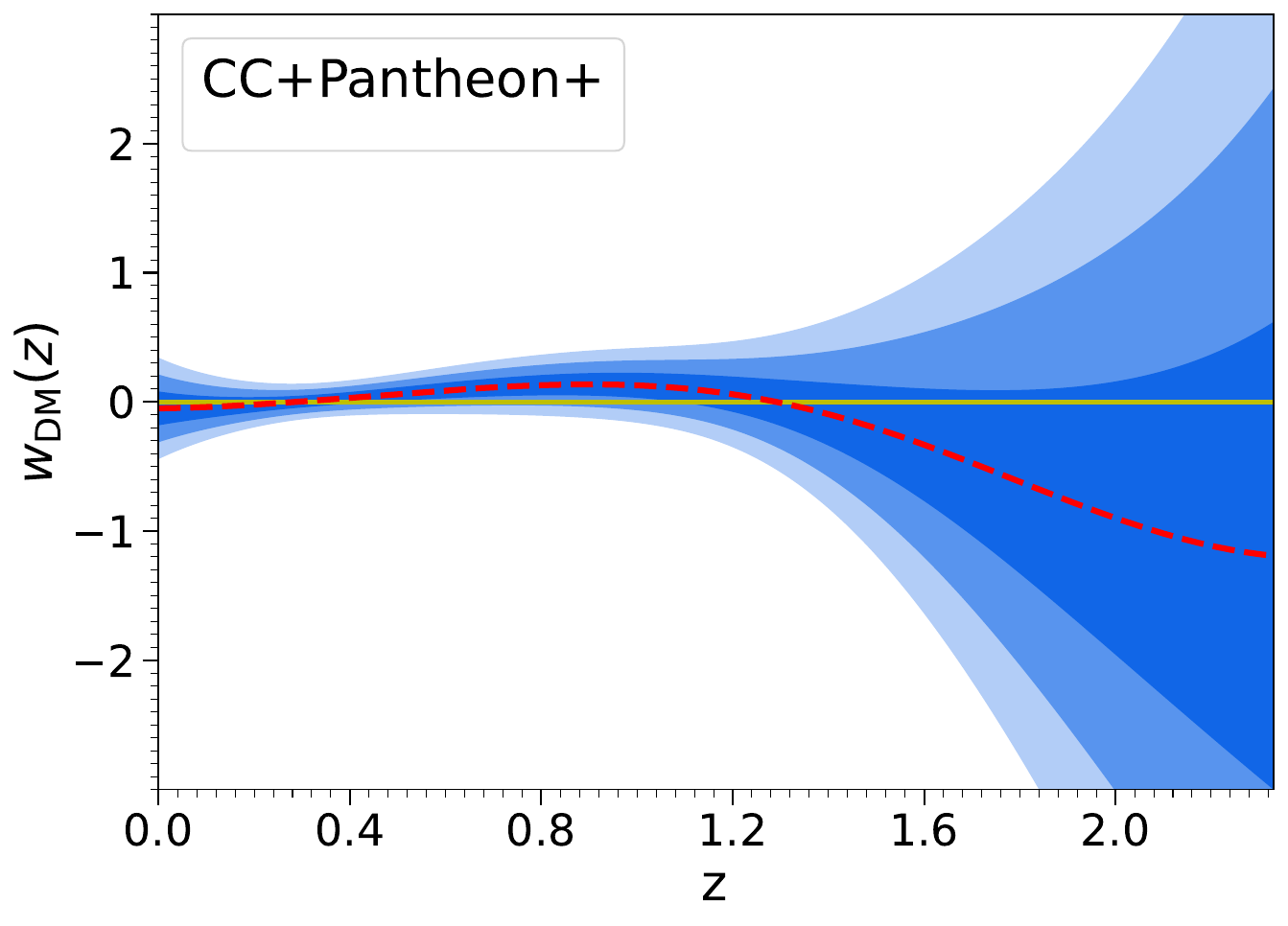}
     }    
    \makebox[10cm][c]{
     \includegraphics[trim = 0mm  0mm 0mm 0mm, clip, width=5.6cm, height=4.cm]{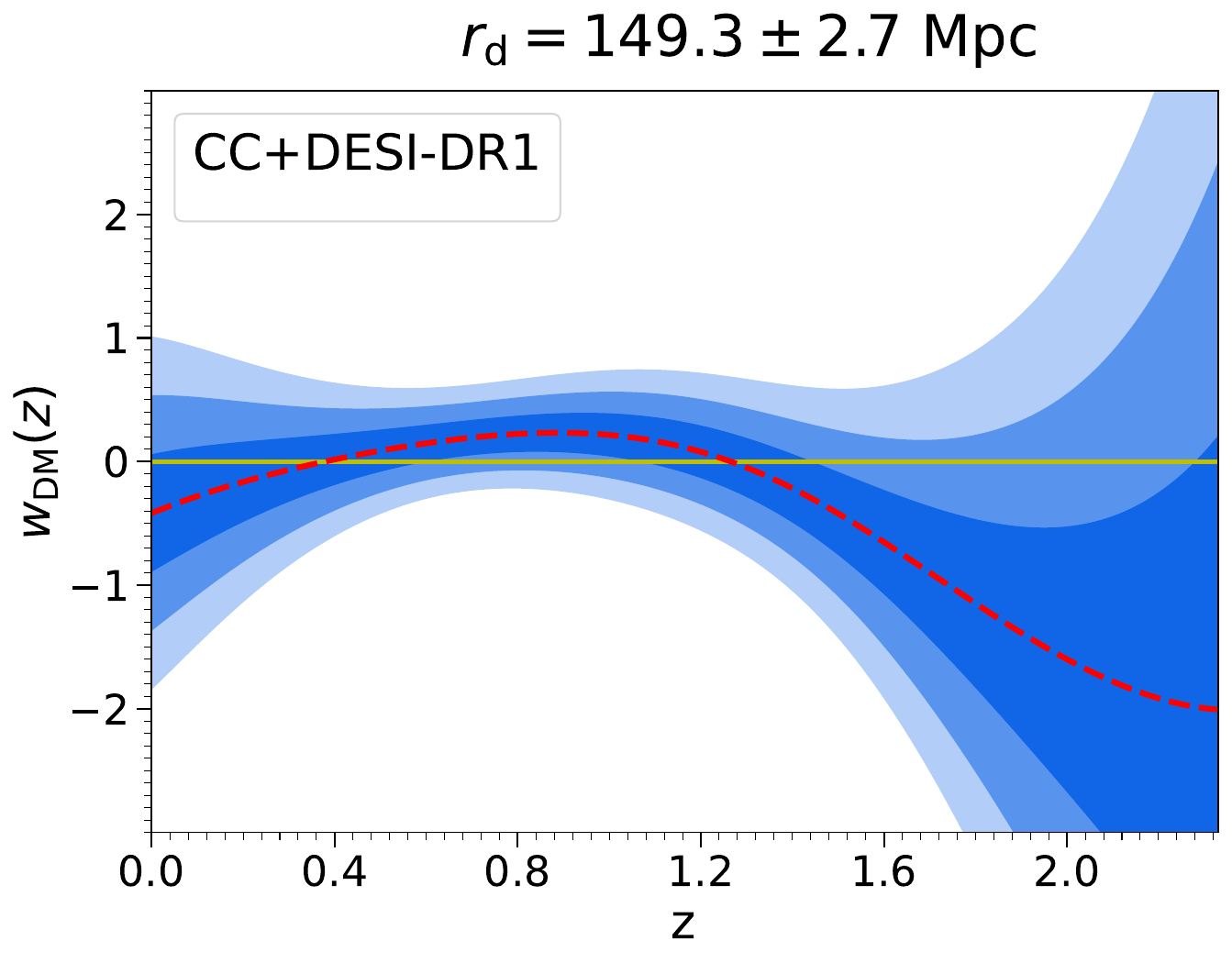}
     \includegraphics[trim = 0mm  0mm 0mm 0mm, clip, width=5.6cm, height=4.cm]{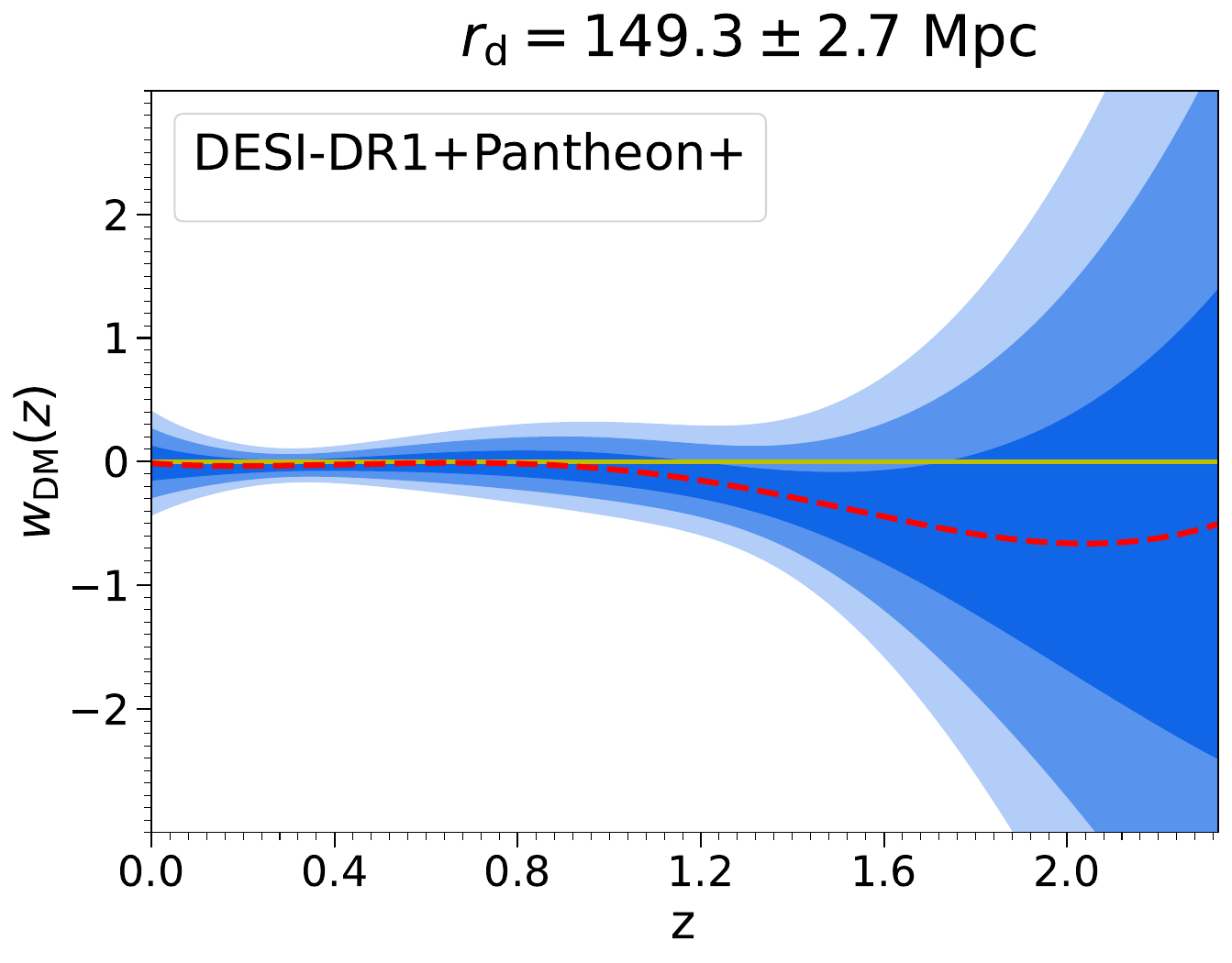}
     \includegraphics[trim = 0mm  0mm 0mm 0mm, clip, width=5.6cm, height=4.cm]{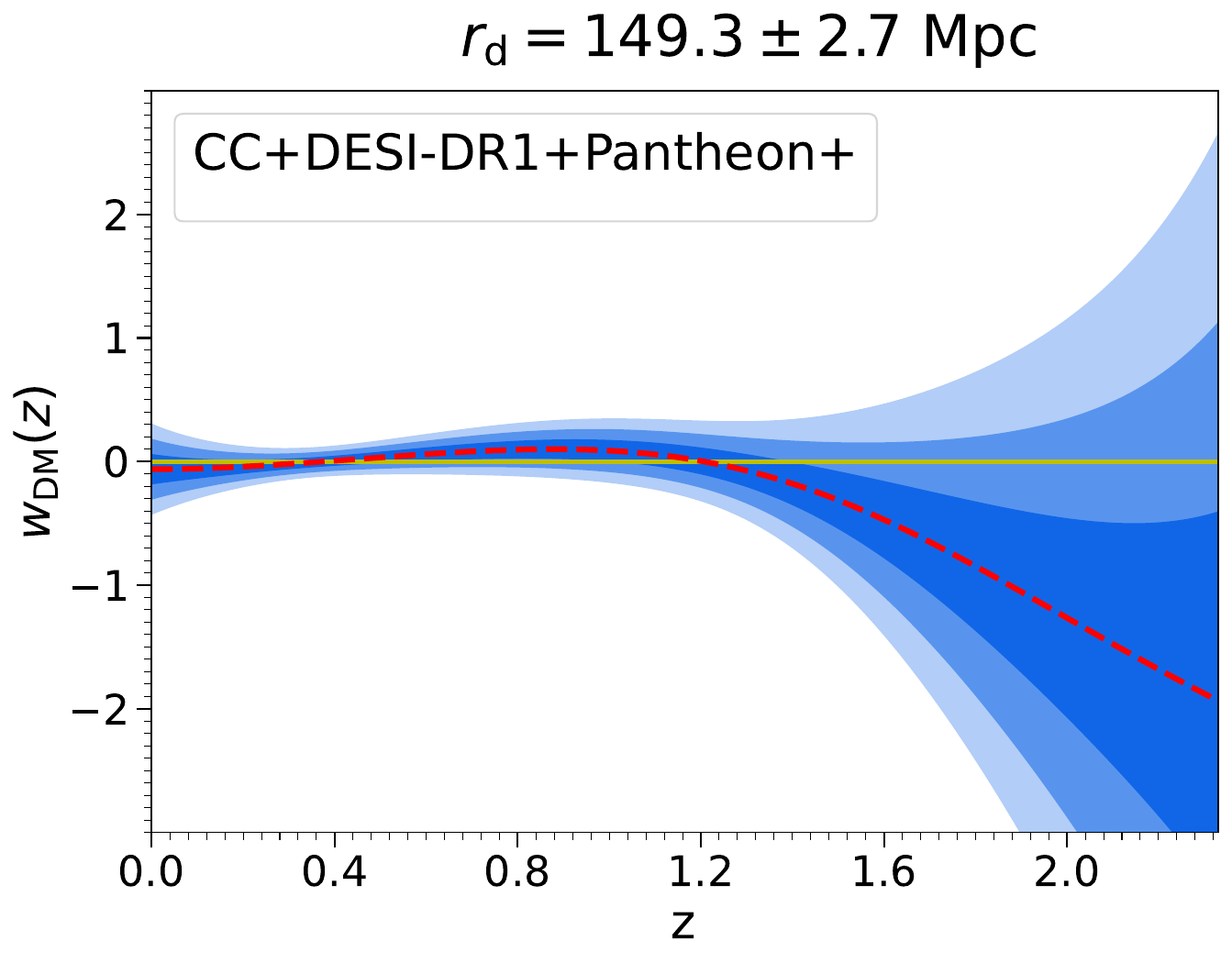}
     }
     \makebox[10cm][c]{
     \includegraphics[trim = 0mm  0mm 0mm 0mm, clip,width=5.6cm, height=4.cm]{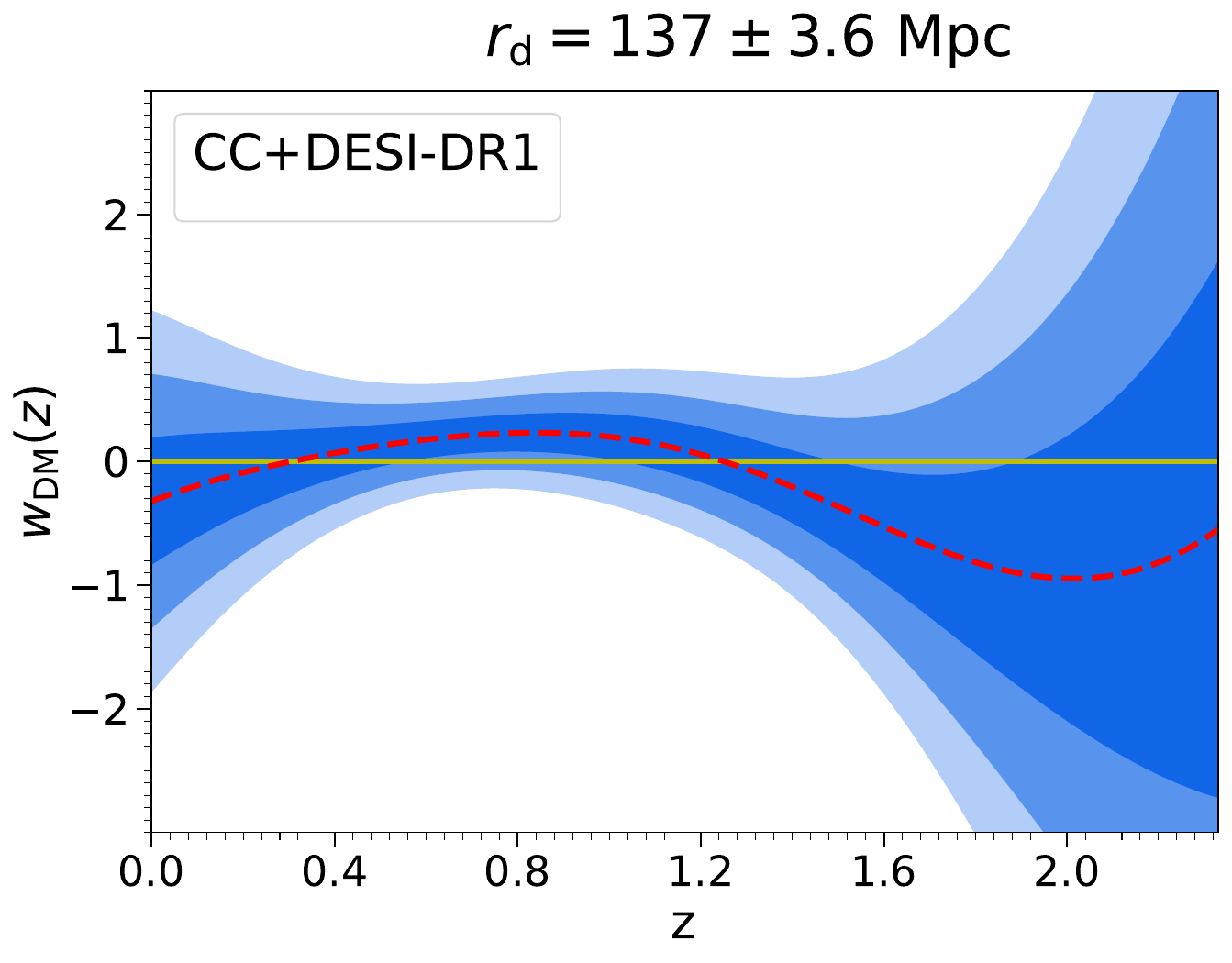}
     \includegraphics[trim = 0mm  0mm 0mm 0mm, clip, width=5.6cm, height=4.cm]{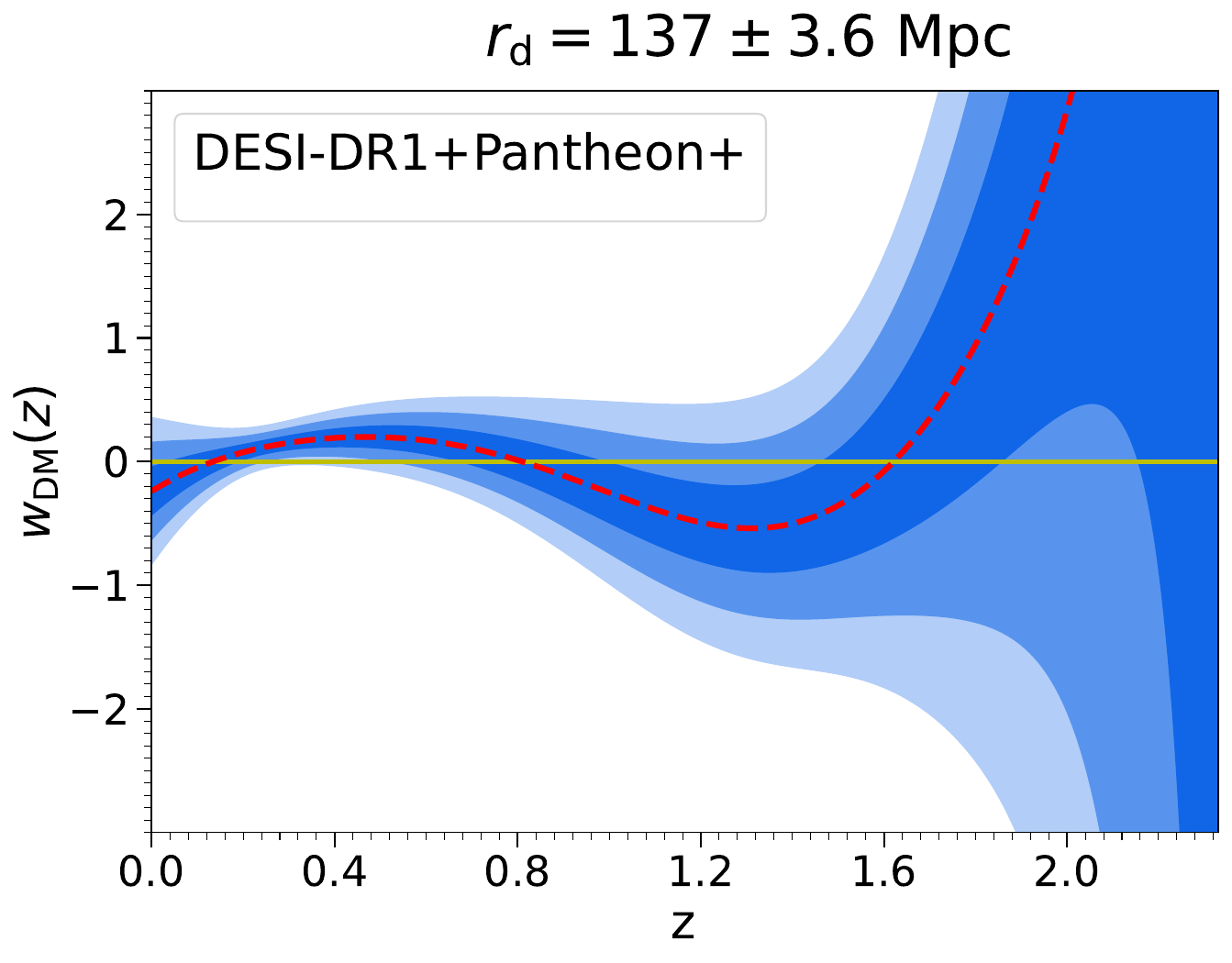}
     \includegraphics[trim = 0mm  0mm 0mm 0mm, clip, width=5.6cm, height=4.cm]{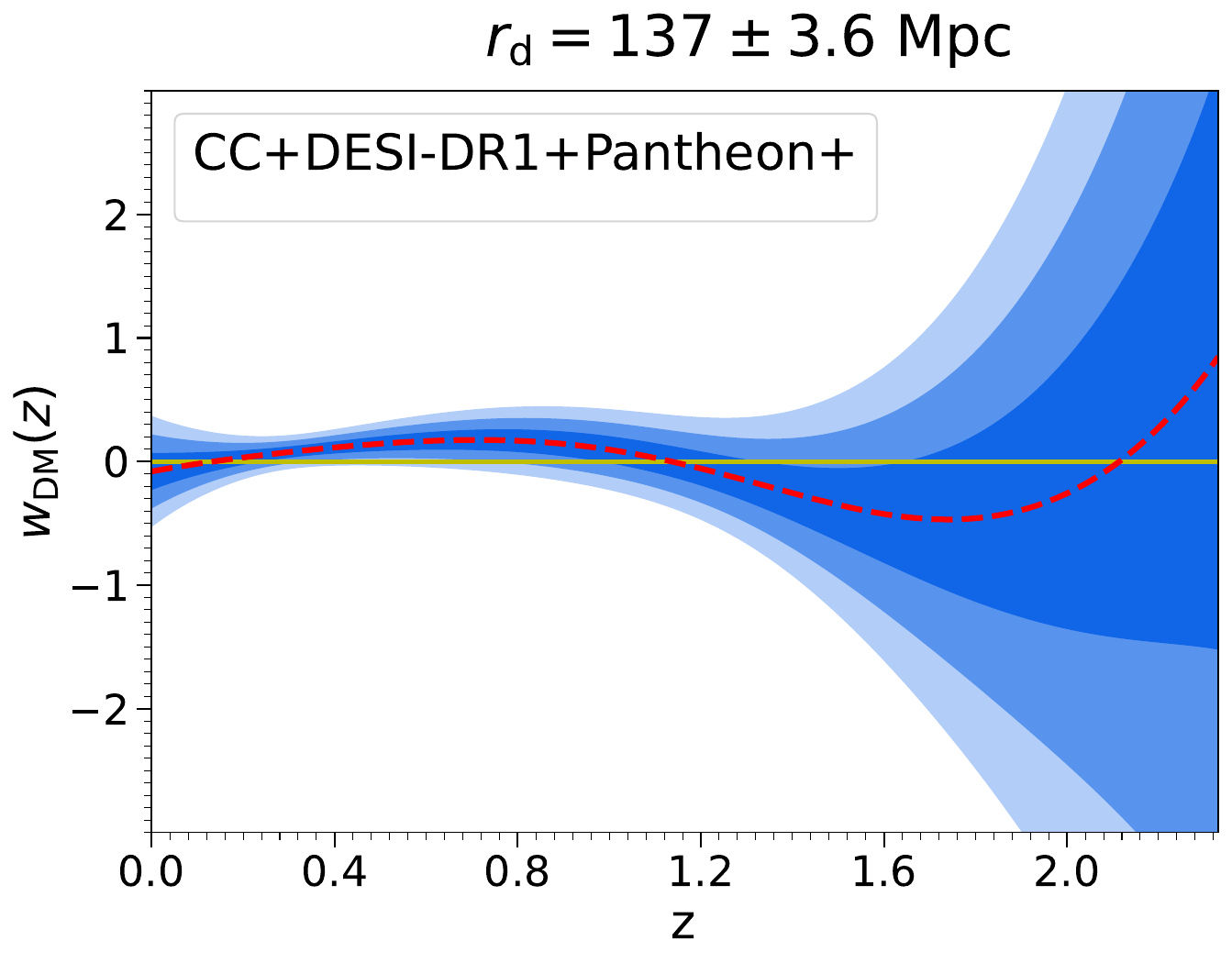}
     }
     \makebox[10cm][c]{
     \includegraphics[trim = 0mm  0mm 0mm 0mm, clip, width=5.6cm, height=4.cm]{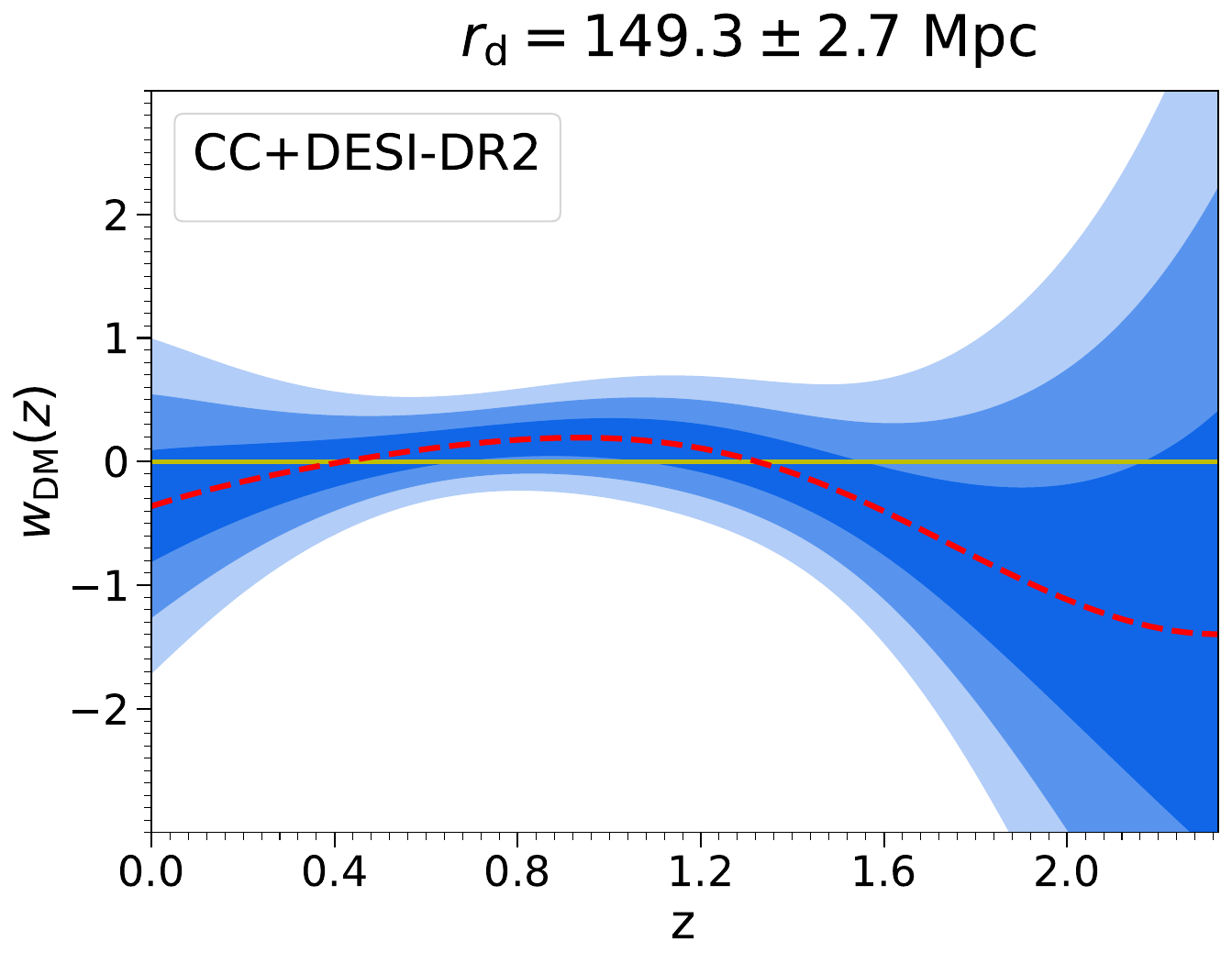}
     \includegraphics[trim = 0mm  0mm 0mm 0mm, clip,width=5.6cm, height=4.cm]{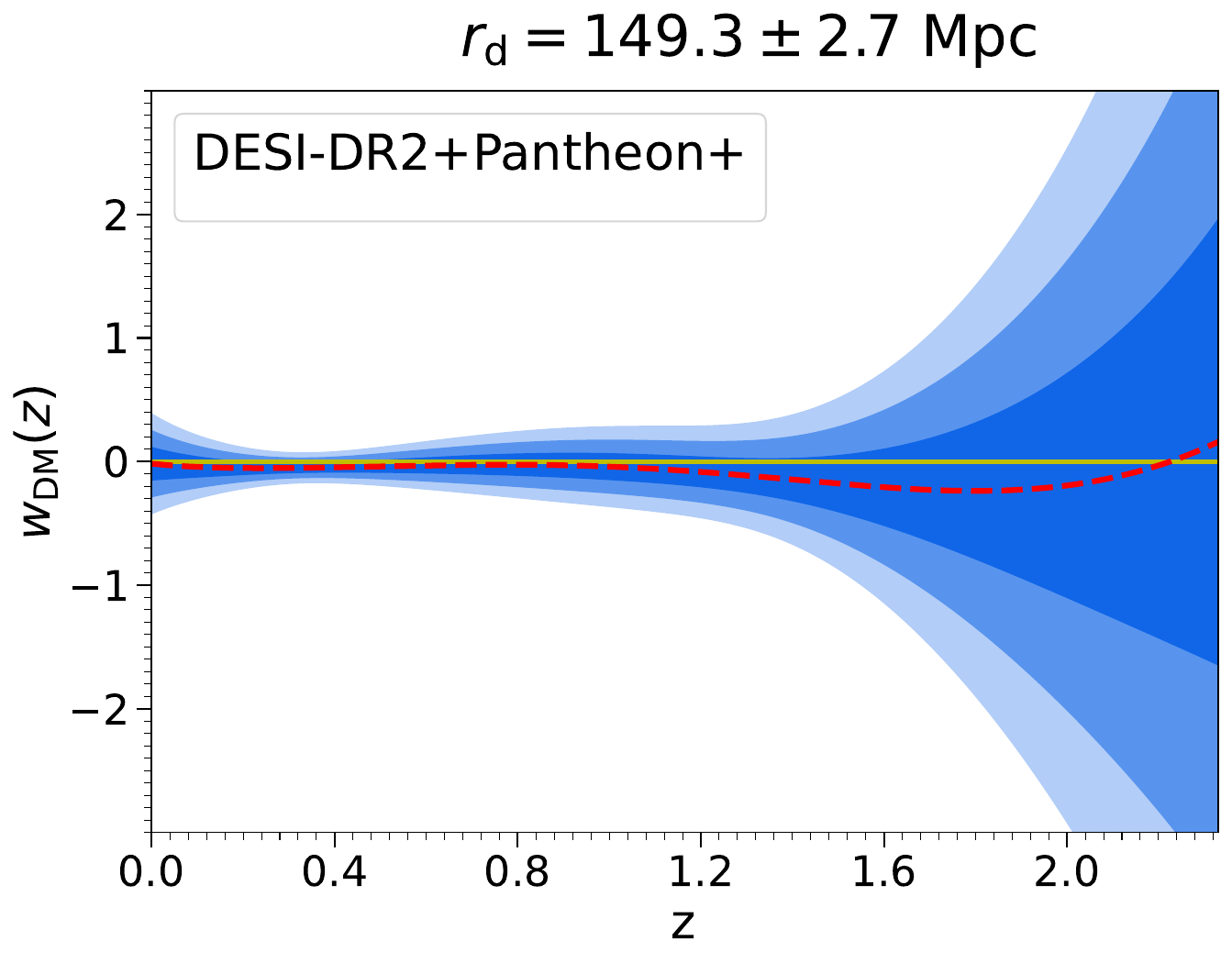}
     \includegraphics[trim = 0mm  0mm 0mm 0mm, clip, width=5.6cm, height=4.cm]{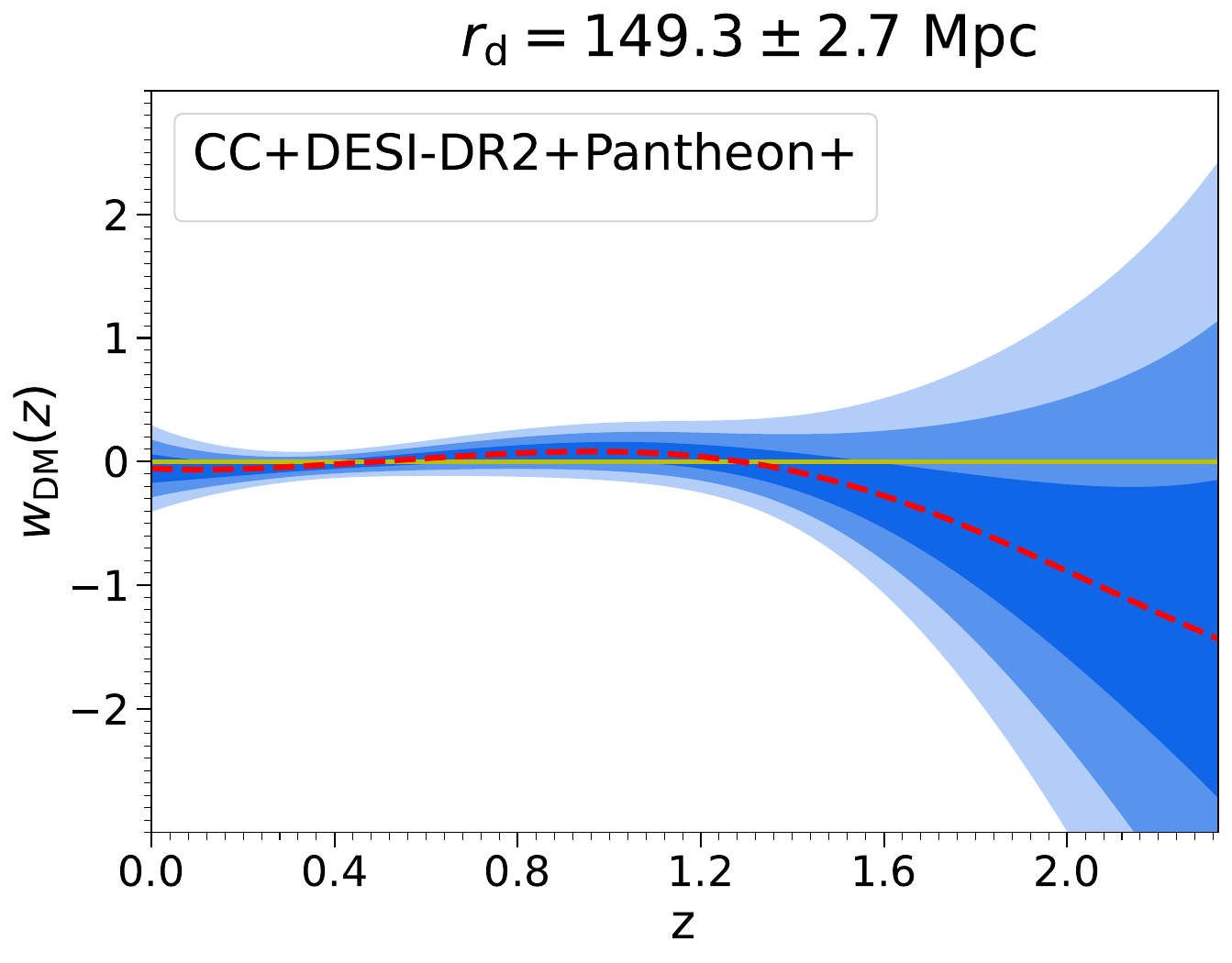}
     } 
     \makebox[10cm][c]{
     \includegraphics[trim = 0mm  0mm 0mm 0mm, clip, width=5.6cm, height=4.cm]{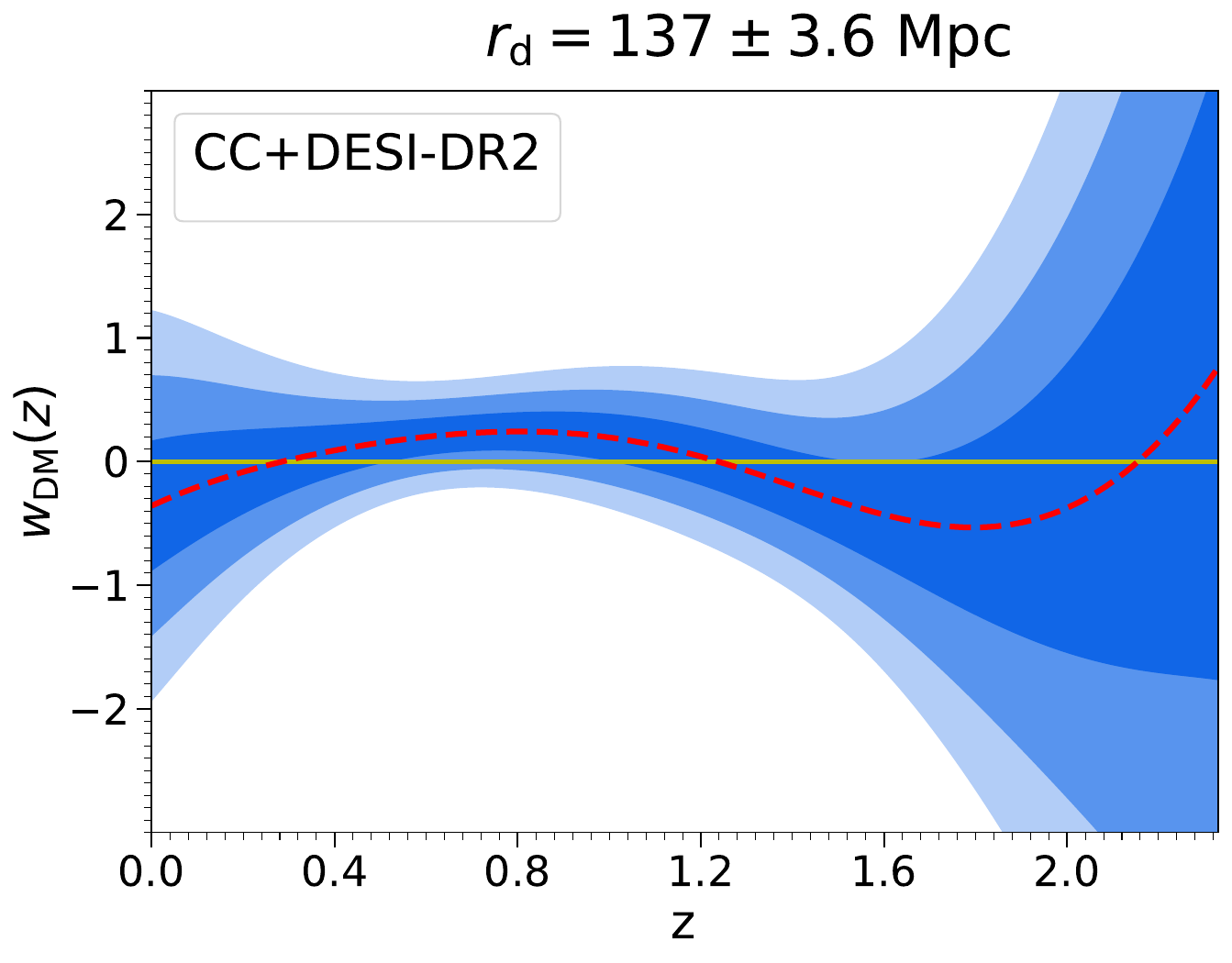}
     \includegraphics[trim = 0mm  0mm 0mm 0mm, clip,width=5.6cm, height=4.cm]{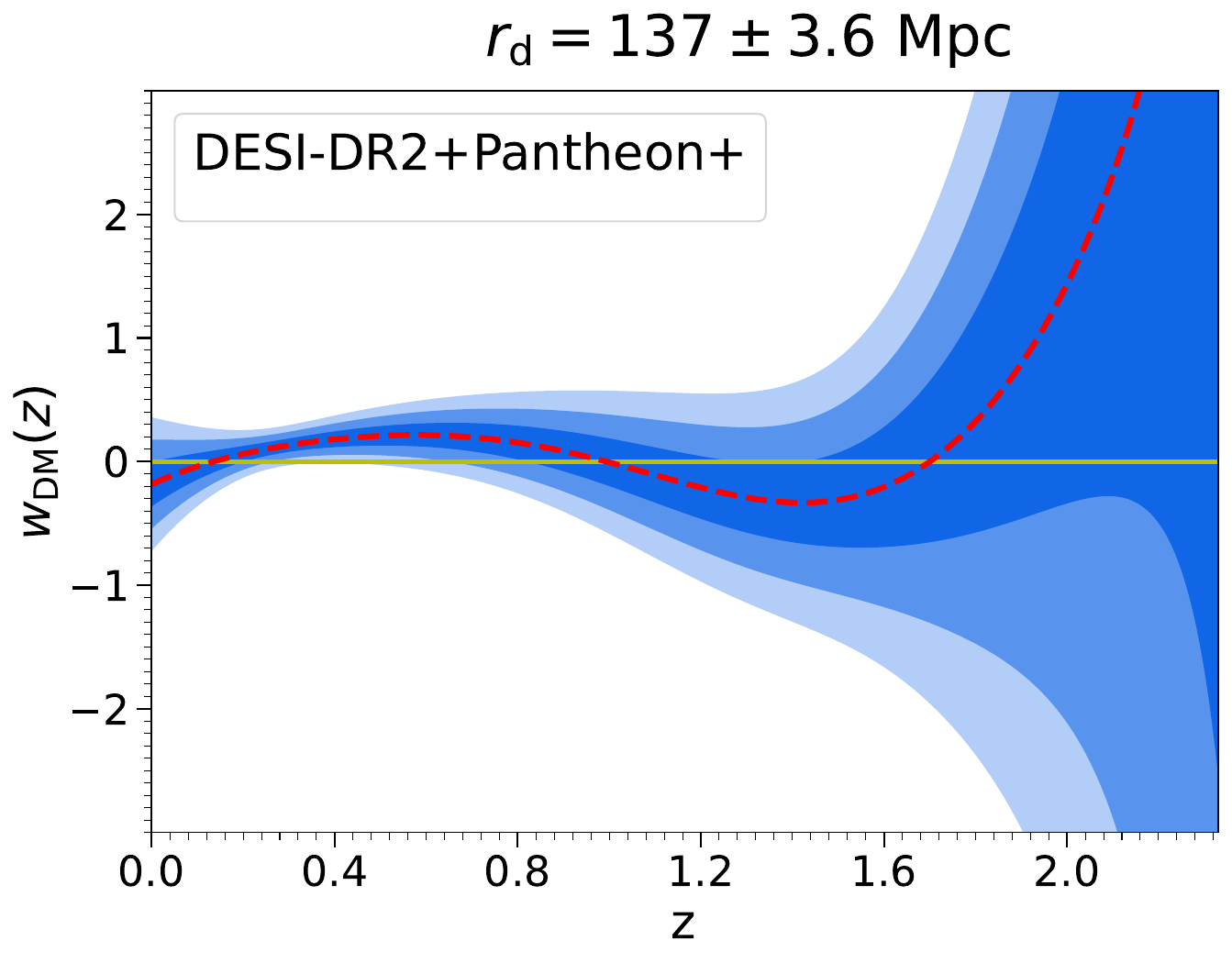}
     \includegraphics[trim = 0mm  0mm 0mm 0mm, clip, width=5.6cm, height=4.cm]{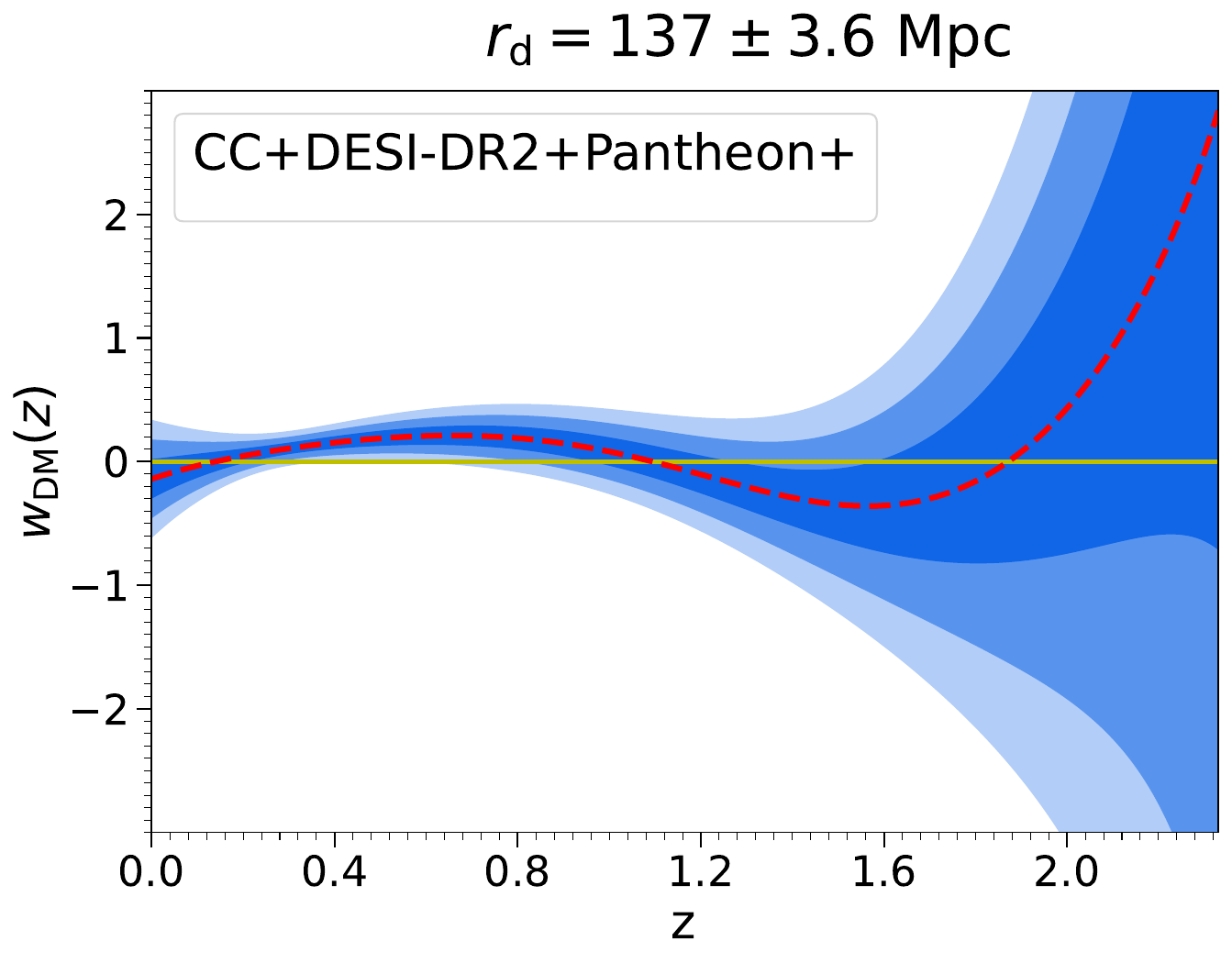}
     } 
\caption{Reconstruction of $w_{\rm DM} (z)$ using the {\it squared exponential kernel} of the Gaussian approach, under the minimal assumption that DE  corresponds to the cosmological constant, i.e., $w_{\rm DE} = -1$, for the CC, Pantheon+, CC+Pantheon+, DESI+Pantheon+, and CC+DESI+Pantheon+ datasets. The red dashed curve represents the mean curve of the reconstructed $w_{\rm DM} (z)$, while the solid horizontal line  corresponds to $w_{\rm DM} (z) = 0$.} 
\label{fig:wdm_gpr-kernel}
\end{figure*}

\begin{figure}
\includegraphics[width=0.5\textwidth]{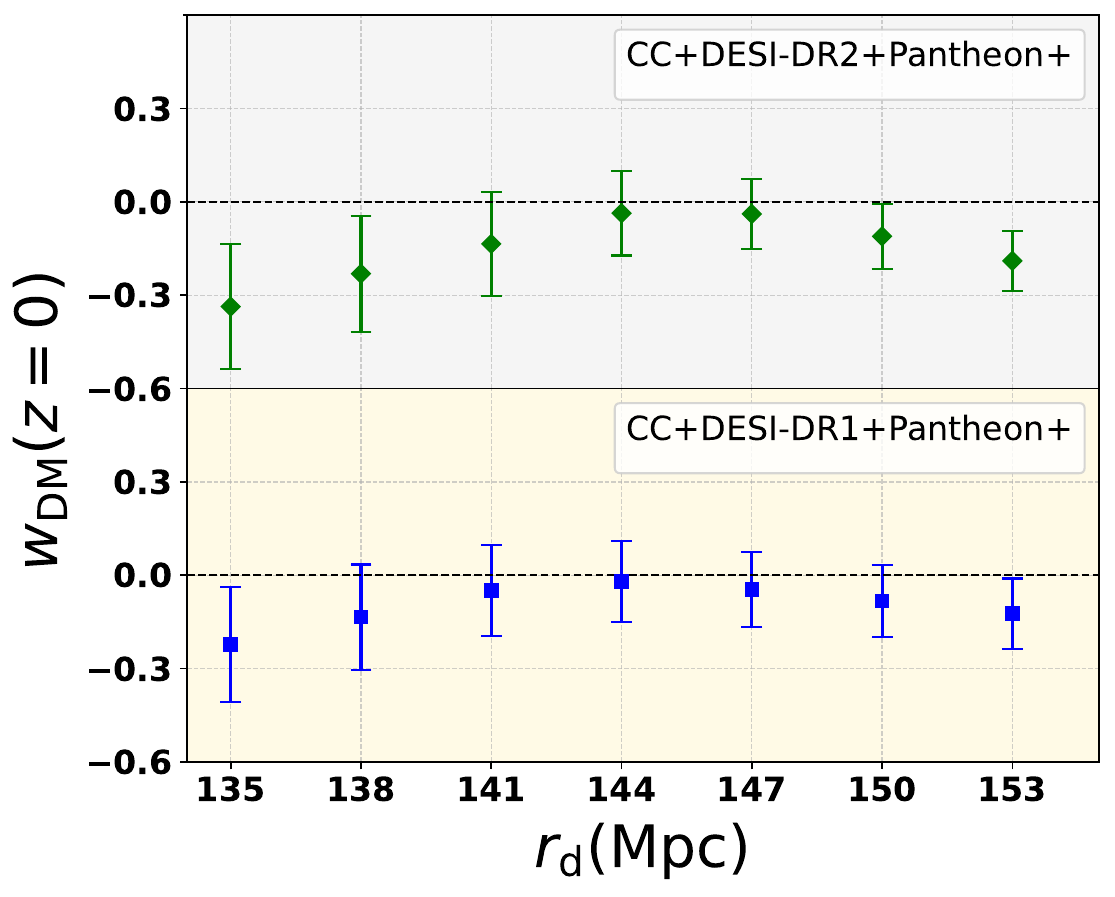}
\caption{Whisker plot showing the 68\%~CL constraints on the present-day value of $w_{\rm DM}(z)$, i.e., $w_{\rm DM}(z = 0)$, for different discrete values of $r_{\rm d}$ (in Mpc) within the interval $135~\mathrm{Mpc} \leq r_{\rm d} \leq 153~\mathrm{Mpc}$, obtained using the {\it squared exponential} kernel of the Gaussian process and two combined datasets: CC+DESI-DR2+Pantheon+ (upper panel) and CC+DESI-DR1+Pantheon+ (lower panel).}
    \label{fig:whisker-GPR}
\end{figure} 
\begin{figure*}[ht]
    \centering
    \makebox[10cm][c]{%
        \includegraphics[width=0.49\textwidth, height=3.5cm]{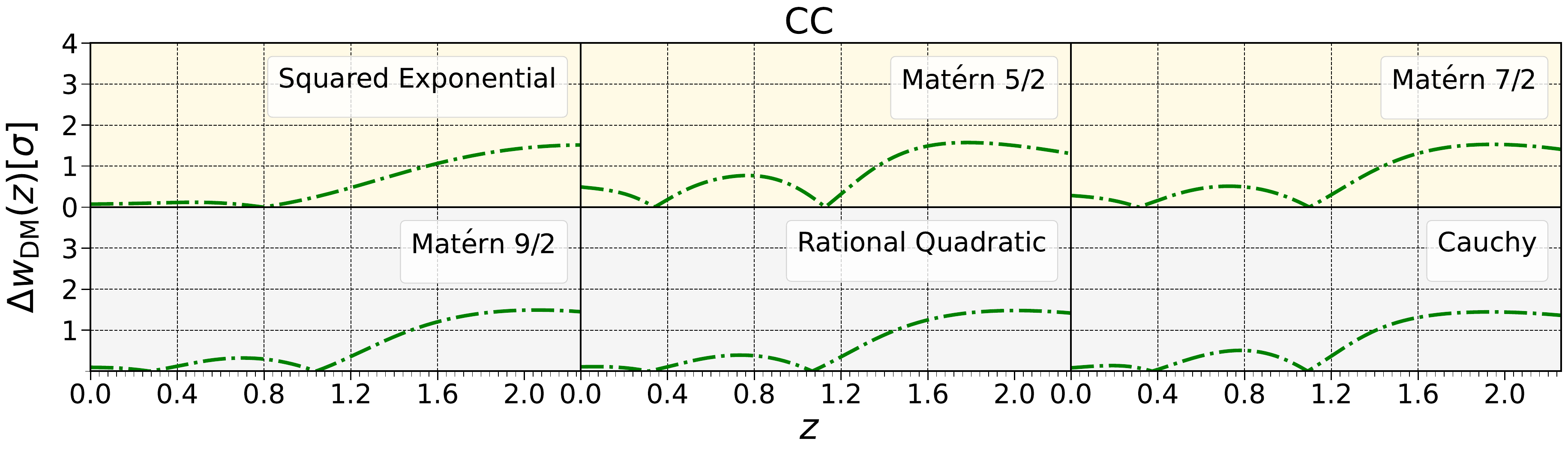}%
    \includegraphics[width=0.49\textwidth, height=3.5cm]{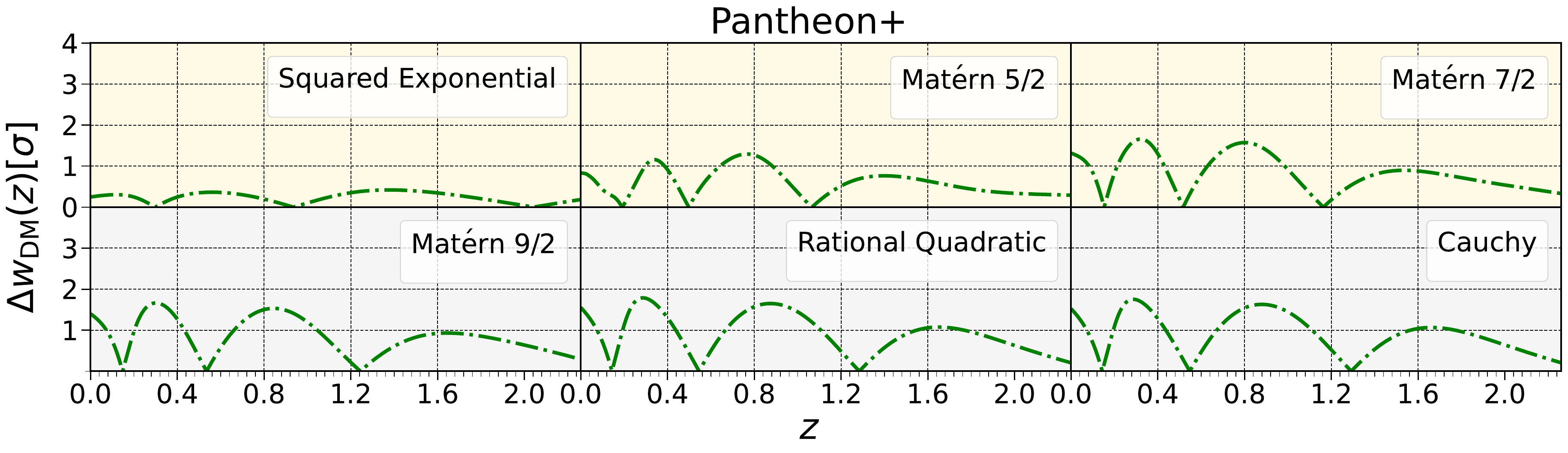}%
    }
    \makebox[10cm][c]{%
        \includegraphics[width=0.49\textwidth, height=3.5cm]{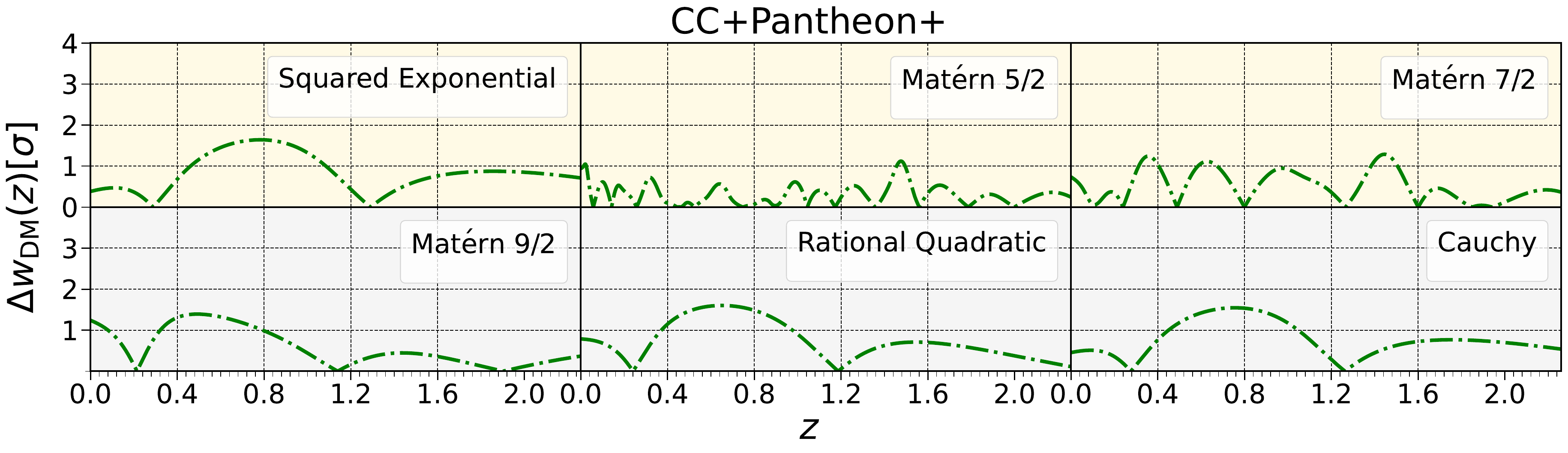}%
    \includegraphics[width=0.49\textwidth, height=3.5cm]{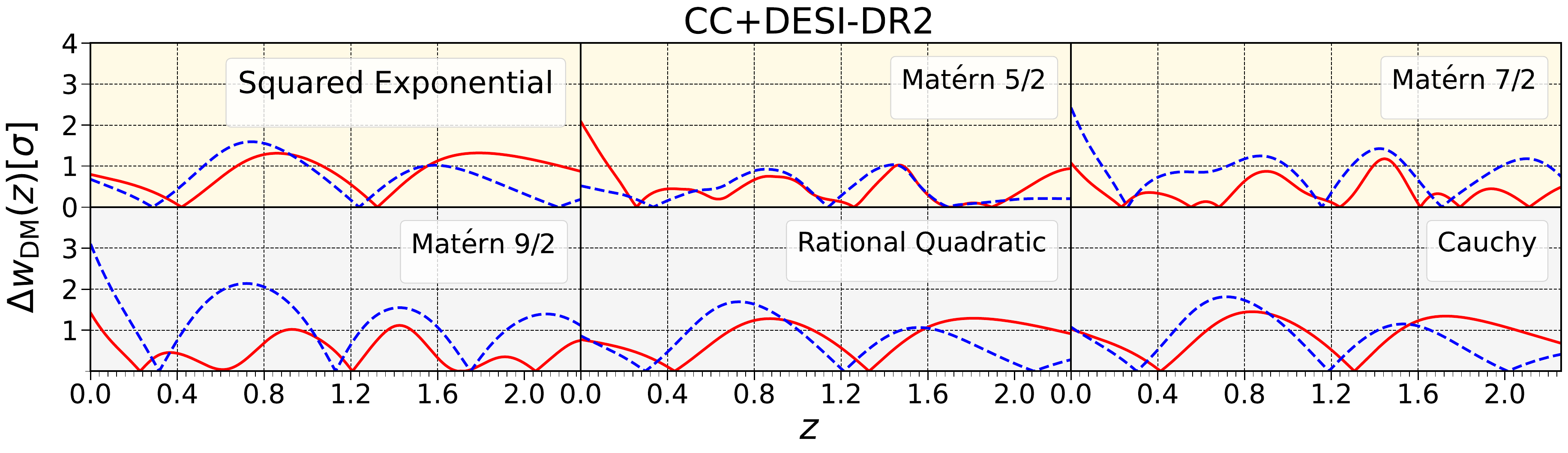}%
        
    }

\makebox[10cm][c]{%
        \includegraphics[width=0.49\textwidth, height=3.5cm]{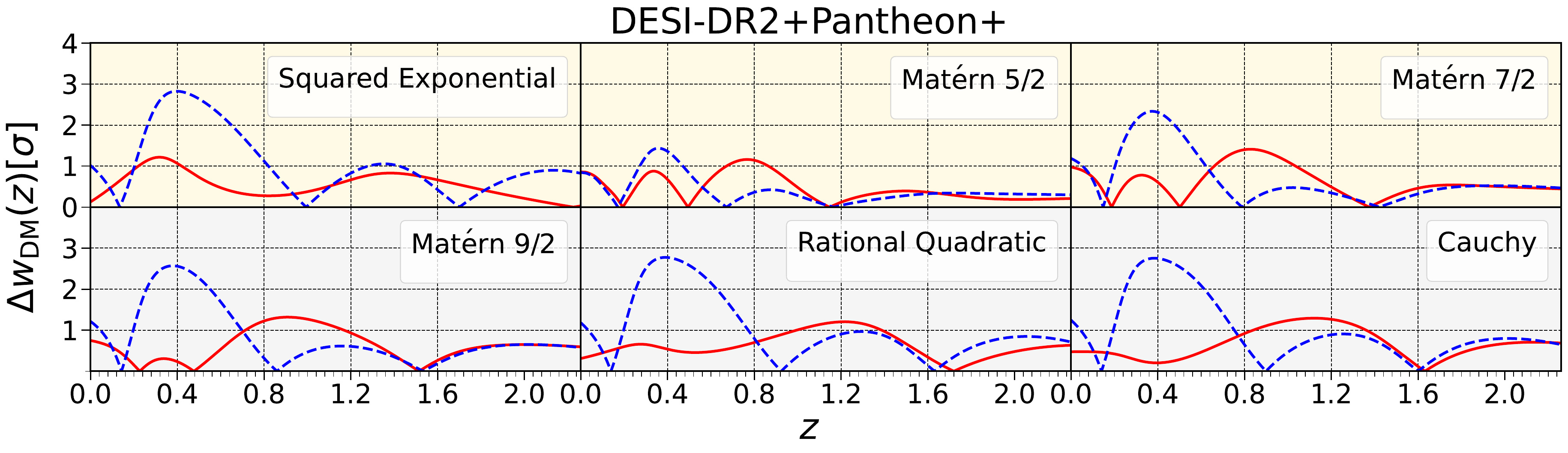}%
        \includegraphics[width=0.49\textwidth, height=3.5cm]{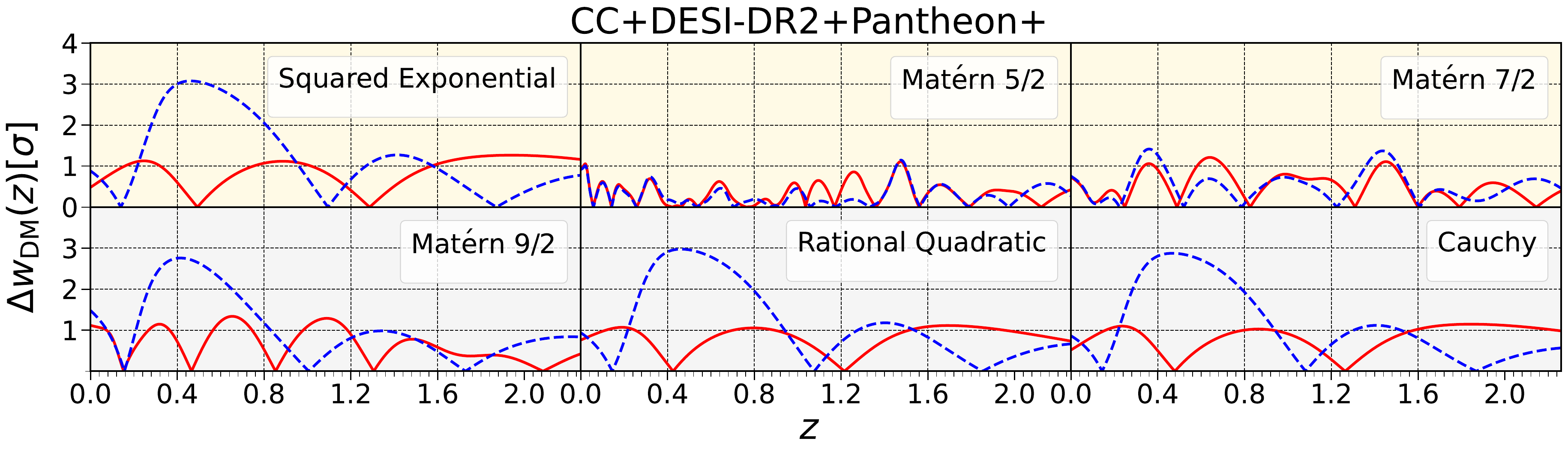}%
    }
\caption{Deviation of $w_{\rm DM}(z)$ from zero, quantified by $\Delta w_{\rm DM} = |w_{\rm DM}(z) - 0| / |\sigma_{w_{\rm DM}(z)}|$, for different datasets across various kernels. For the datasets where DESI DR2 BAO is included, the solid red line corresponds to $r_{\rm d} = 149.3 \pm 2.7$ Mpc, while the dashed blue line corresponds to $r_{\rm d} = 137 \pm 3.6$ Mpc.  Non-zero evidence for $w_{\rm DM}(z)$ across the redshift range is observed, regardless of the choice of kernel.} 
\label{fig:deviation-various-kernels}
\end{figure*}
\begin{figure*}[t!]
    \centering
    \makebox[9.5cm][c]{
     \includegraphics[trim = 5mm  0mm 10mm 0mm, clip, width=6.5cm, height=5.cm]{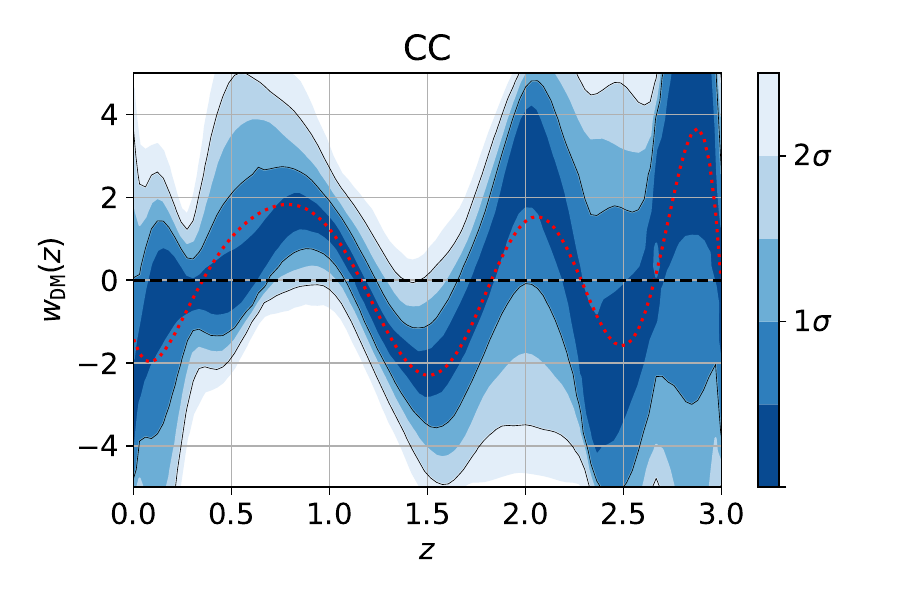}
     \includegraphics[trim = 5mm  0mm 10mm 0mm, clip, width=6.5cm, height=5.cm]{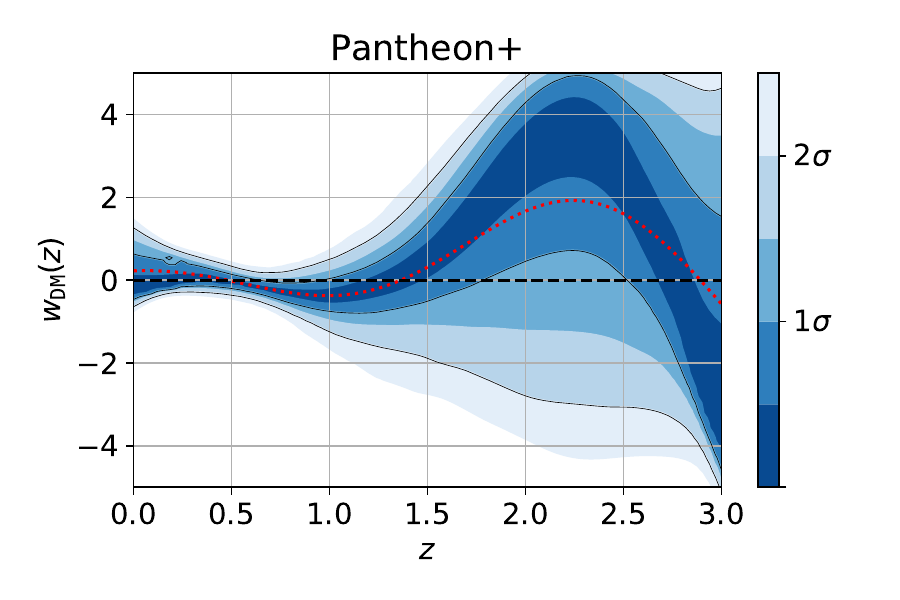}
     \includegraphics[trim = 5mm  0mm 10mm 0mm, clip, width=6.5cm, height=5.cm]{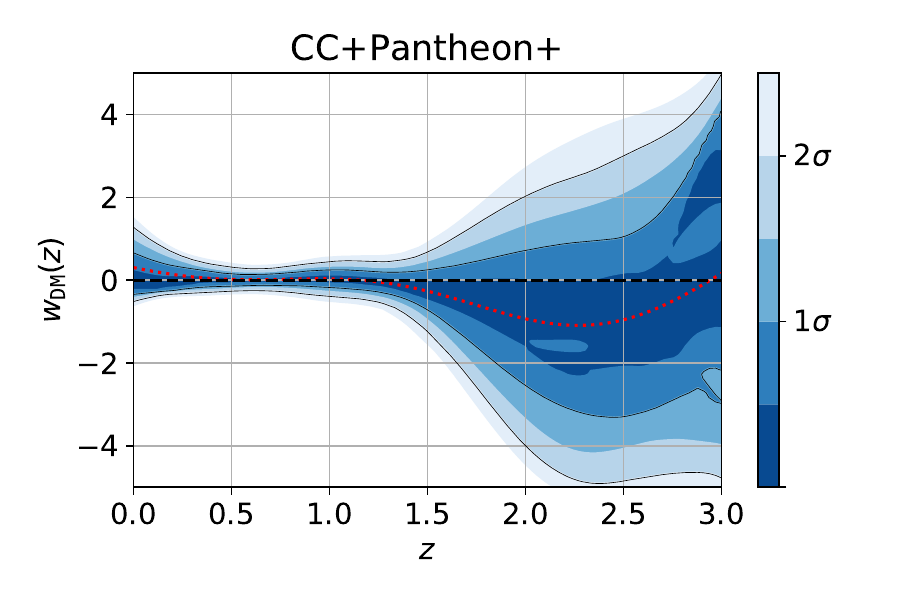}
     }
    \makebox[9.5cm][c]{
     \includegraphics[trim = 5mm  0mm 10mm 0mm, clip, width=6.5cm, height=5.cm]{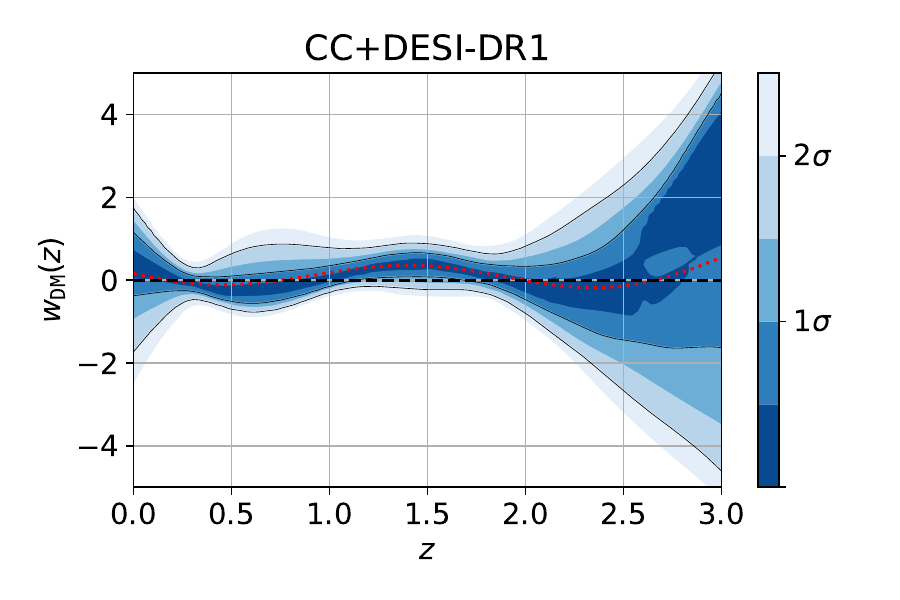}
     \includegraphics[trim = 5mm  0mm 10mm 0mm, clip, width=6.5cm, height=5.cm]{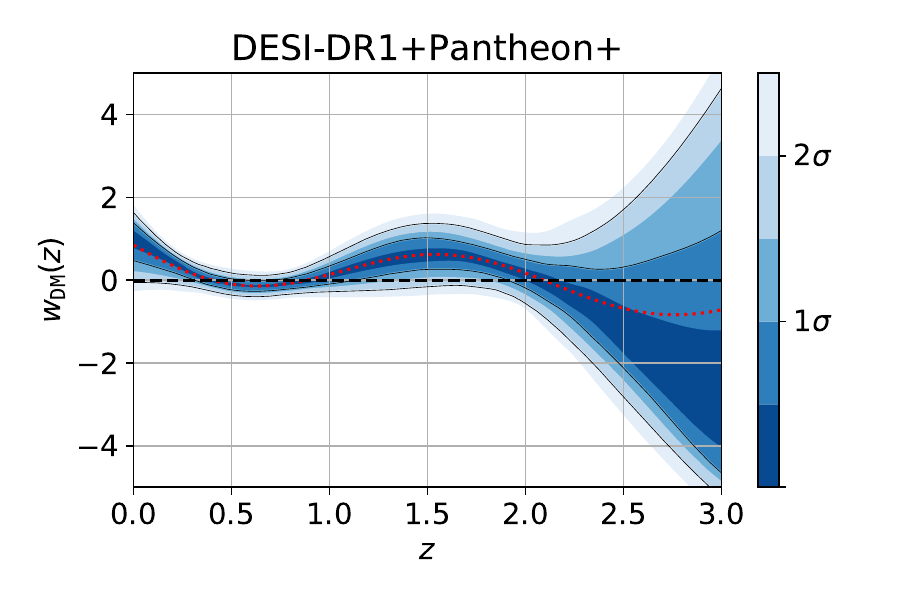}
     \includegraphics[trim = 5mm  0mm 10mm 0mm, clip, width=6.5cm, height=5.cm]{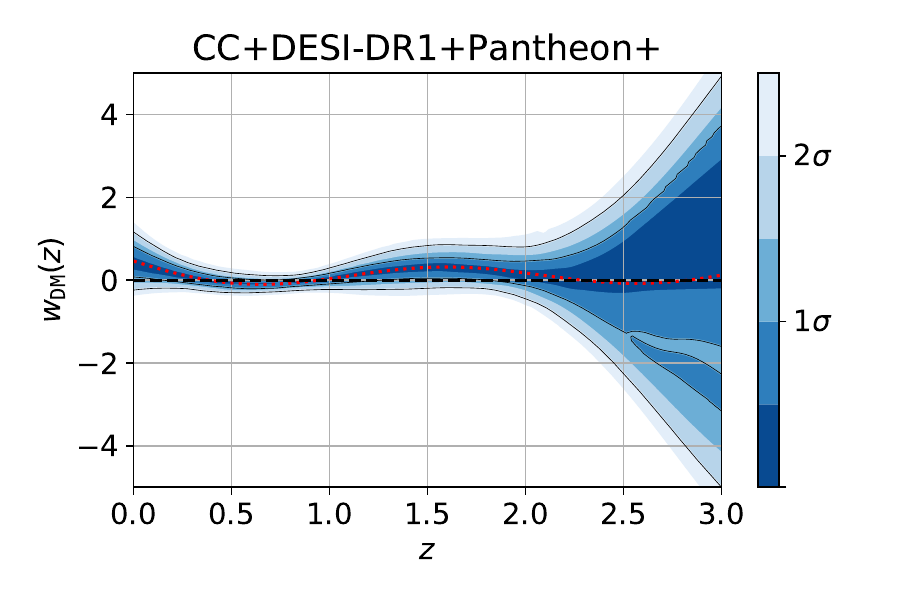}
     }    
     \makebox[9.5cm][c]{
     \includegraphics[trim = 5mm  0mm 10mm 0mm, clip, width=6.5cm, height=5.cm]{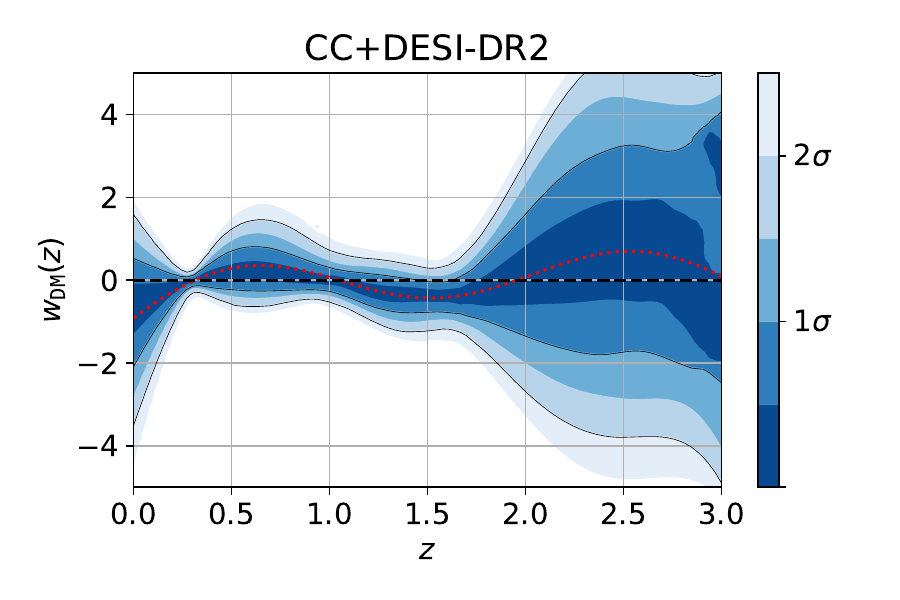}
     \includegraphics[trim = 5mm  0mm 10mm 0mm, clip, width=6.5cm, height=5.cm]{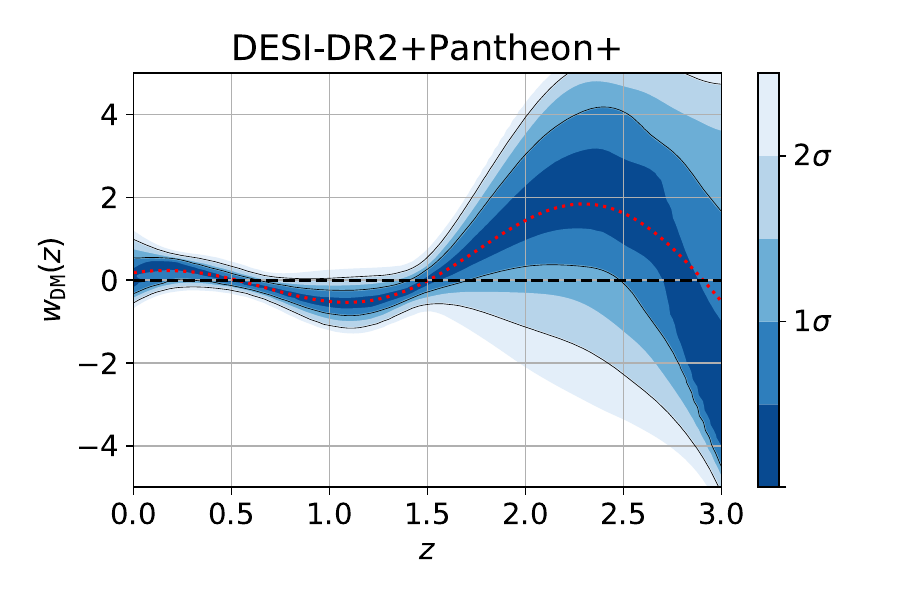}
     \includegraphics[trim = 5mm  0mm 10mm 0mm, clip, width=6.5cm, height=5.cm]{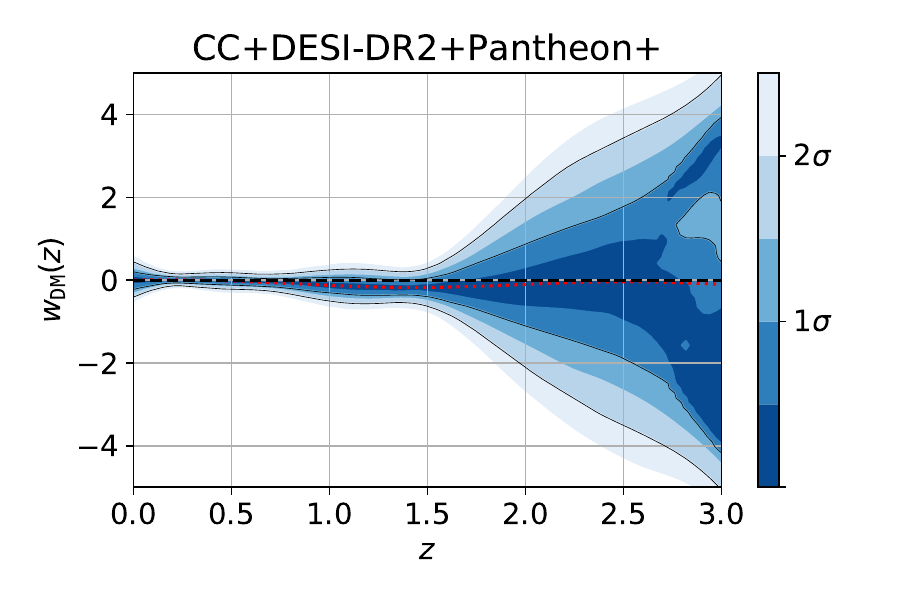}
     } 
\caption{Parametric reconstructions of $w_{\rm DM}$ using different datasets, considering the binned Gaussian approach. These plots represent the functional posterior and were generated using the publicly available Python library \texttt{fgivenx}~\cite{handley2019fgivenx}. The red dotted curve represents the best-fit curve for each case and the horizontal dotted line corresponds to $w_{\rm DM} =0$. }
\label{fig:wdm_gp}
\end{figure*}

\subsection{Kernel approach}

\subsubsection{CC and CC+DESI}
\label{sec-kernel--cc-cc+desi}

We begin with the results for CC alone, considering the \textit{squared exponential} kernel, and later summarize the key findings for the remaining kernels. Using the 31 CC data points listed in Table~\ref{tab:Hz}, we first reconstruct $H(z)$ (upper plot of Fig.~\ref{fig:Hz-dHz}) and $H'(z)$ (lower plot of Fig.~\ref{fig:Hz-dHz}). 
With these reconstructions, one can obtain the dimensionless quantities $E(z)$ and $E'(z)$, which are required to recover Eq.~(\ref{eos-DM-rec-E}). It is important to note that to reconstruct $E(z)$ and $E'(z)$, specifying a value for $H_0$ is required. In principle, any specific value of $H_0$ can be chosen, but different values may impact the final results~\cite{Wang:2020dbt}. Thus, the choice of $H_0$ may, in principle, influence the results, and in general, there is no guiding principle supporting a particular choice of $H_0$. 
However, in this case, we have the possibility of avoiding the arbitrary selection of $H_0$ given that, from the reconstructed graph of $H(z)$ (Fig.~\ref{fig:Hz-dHz}), one can determine the value of $H(z)$ at $z = 0$ by extrapolation. This yields $H_0 = 67.45 \pm 4.75$ km/s/Mpc (see Table~\ref{tab:rec_H0}), which we have used to obtain $E(z)$ and $E'(z)$. 
With these reconstructed functions, we can now reconstruct  $w_{\rm DM}(z)$ using Eq.~(\ref{eos-DM-rec-E}). The topmost plot in the left column of Fig.~\ref{fig:wdm_gpr-kernel} corresponds to the case of CC data alone. From this, we observe that $w_{\rm DM} (z)$ is consistent with zero at low redshifts. Therefore, CC data alone does not indicate any deviation from CDM.

We now present the results using the BAO measurements, namely, $D_{\rm M}/r_{\rm d}$ from DESI DR1 and DESI DR2, combined with the CC dataset. As already argued, reconstructions using BAO data require a fixed value of the $r_{\rm d}$,
which inevitably influences the results. For the CC+DESI data combination (and for all combinations considered hereafter where DESI is used), we adopt two distinct values of $r_{\rm d}$: namely, $r_{\rm d} = 149.3 \pm 2.7$~Mpc at 68\%~CL~\cite{eBOSS:2020yzd} and $r_{\rm d} = 137 \pm 3.6$~Mpc at 68\%~CL~\cite{Arendse:2019hev}, and perform the reconstructions of $w_{\rm DM}(z)$. 
These two choices are motivated by existing literature: the higher value of $r_{\rm d}$ has been reported by 
Planck 2018 ($r_{\rm d} = 147.09 \pm 0.26$~Mpc at 68\%~CL, Planck TT,TE,EE+lowE+lensing~\cite{Planck:2018vyg}) and by BAO measurements from the Sloan Digital Sky Survey combined with Big Bang Nucleosynthesis ($r_{\rm d} = 149.3 \pm 2.7$~Mpc at 68\%~CL)~\cite{eBOSS:2020yzd}, 
while a nearly model-independent approach based on low-redshift measurements yields a smaller value, $r_{\rm d} = 137 \pm 3.6$~Mpc~\cite{Arendse:2019hev}. 
According to the literature, there is currently no compelling evidence favoring either of these two values. Therefore, for completeness, we consider both a high and a low value of $r_{\rm d}$. 
In this work, we adopt the estimate from~\cite{eBOSS:2020yzd} (i.e., $r_{\rm d} = 149.3 \pm 2.7$~Mpc at 68\%~CL), since it is consistent with the Planck 2018 value within the same confidence level. The uncertainties in $r_{\rm d}$ for these two measurements are taken into account by applying the error propagation rule to derive the covariance matrix of $D_{\rm M}$ and to study how variations in $r_{\rm d}$ affect the reconstructed $w_{\rm DM}(z)$ and its associated uncertainty.\footnote{We note that for Fig.~\ref{fig:whisker-GPR}, where we calculate $w_{\rm DM}(z=0)$ for different $r_{\rm d}$ values, we use a few discrete $r_{\rm d}$ values solely to illustrate the trend in $w_{\rm DM}(z=0)$.}
The left column (excluding the upper plot) of Fig.~\ref{fig:wdm_gpr-kernel} presents the reconstructed graphs of $w_{\rm DM}$ for CC+DESI-DR1 and CC+DESI-DR2. Focusing on CC+DESI-DR1 for both choices of $r_{\rm d}$, we notice mild preference for $w_{\rm DM} < 0$ (at slightly more than 68\% CL) in the high-redshift regime\footnote{One can further notice that with the increase of $r_{\rm d}$, mild preference of a non-null nature of $w_{\rm DM}$ is pronounced only in the high-redshift regime. It is important to note, however, that beyond $z \sim 2$, the amount of data, and consequently its constraining power, is limited.} and of $w_{\rm DM} > 0$ in the intermediate-redshift regime. However, in the low-redshift regime, $w_{\rm DM}$ is consistent with zero. 
This is particularly interesting because, although the hint for $w_{\rm DM} \neq 0$ is statistically mild according to these datasets, the indication of a sign-changing behavior in $w_{\rm DM}$ warrants further investigation.
When DESI-DR1 is replaced with DESI-DR2 in the combined dataset, and assuming $r_{\rm d} = 149.3 \pm 2.7$ Mpc, the results remain nearly identical to those obtained with CC+DESI-DR1. Specifically, we continue to observe a mild preference for $w_{\rm DM} < 0$ at high redshifts and $w_{\rm DM} > 0$ at intermediate redshifts. However, these indications are slightly weakened when DESI-DR2 is used. Overall, the presence of a non-zero $w_{\rm DM}$ throughout the expansion history of the universe cannot be ruled out.

\subsubsection{Pantheon+ and DESI+Pantheon+}

We now switch our focus to the reconstruction of $w_{\rm DM}$ using Pantheon+ alone for the {\it squared exponential kernel}. Unlike the previous case with CC alone, where $H(z)$ was obtained directly, here we need to reconstruct $D(z)$ since Eq.~\eqref{eos-DM-rec-D} involves it. In order to reconstruct $D(z)$, we need to specify a value of $H_0$, and this can be efficiently obtained through the relation between $D_{\rm M}'(z)$ and $H(z)$. As $D_{\rm M}'(z) = c/H(z)$ and hence $D_{\rm M}'(0) = c/H_0$, the reconstructed value of $H(z)$ at $z = 0$ is $68.74 \pm 0.45$~km/s/Mpc. 
We refer to the middle column of Fig.~\ref{fig:wdm_gpr-kernel}, where the top graph corresponds to the reconstruction using Pantheon+ only. According to the results, Pantheon+ by itself does not suggest any hint for non-cold DM, since we observe that $w_{\rm DM}$ is in agreement with its null value within the 68\% CL.

Now, we consider the combined analyses with DESI+Pantheon+, using both DESI DR1 and DESI DR2. Following the strategy described in Section~\ref{sec-kernel--cc-cc+desi}, we performed the reconstructions of $w_{\rm DM}$ using two values of $r_{\rm d}$: $149.3 \pm 2.7$ Mpc and $137 \pm 3.6$ Mpc. The middle column (excluding the upper plot) of Fig.~\ref{fig:wdm_gpr-kernel} presents the reconstructions of $w_{\rm DM}(z)$ for the combined datasets. 
Starting with DESI-DR1+Pantheon+, we notice that for $r_{\rm d} = 149.3 \pm 2.7$ Mpc, there is a very mild hint of $w_{\rm DM} < 0$ (at slightly more than 68\% CL) in the high-redshift regime, while in the low-redshift regime no hint in favor of $w_{\rm DM} \neq 0$ is found. 
However, for DESI-DR1+Pantheon+ with $r_{\rm d} = 137 \pm 3.6$ Mpc, we observe that $w_{\rm DM}$ shows a preference for positive values (where $w_{\rm DM} > 0$ at slightly more than 68\% CL) in the high-redshift regime ($z \sim 2$), then enters into the negative region where $w_{\rm DM} < 0$ (at slightly more than 68\% CL) as $z$ decreases; after that, it transits  to $w_{\rm DM} > 0$ (at slightly more than 95\% CL) close to the present epoch, and finally at $z = 0$, we find a mild preference of $w_{\rm DM} < 0$ at slightly more than 68\% CL.

When replacing DESI-DR1 with DESI-DR2, we find that—unlike the DESI-DR1+Pantheon+ combination with $r_{\rm d} = 149.3 \pm 2.7$ Mpc, $w_{\rm DM}$ shows no clear preference for a non-zero value throughout the expansion history of the universe, and CDM  remains favored in this case.
However, when using $r_{\rm d} = 137 \pm 3.6$ Mpc, the results exhibit some notable changes. While the  hint for a non-zero $w_{\rm DM}$ at high redshift vanishes, a preference for $w_{\rm DM} > 0$ near the present epoch ($z \sim 0.6$) is found at slightly more than 95\% confidence level. Additionally, at $z = 0$, we find $w_{\rm DM} < 0$ at slightly more than 68\% confidence level.

\subsubsection{CC+Pantheon+ and CC+DESI+Pantheon+} 

We now present the results after adding CC to Pantheon+, using the {\it squared exponential} kernel for the GPR. We refer to the right column of Fig.~\ref{fig:wdm_gpr-kernel}, where the top graph in this series corresponds to this reconstruction. In this case, we do not find any compelling evidence of a non-null $w_{\rm DM}(z)$ during the expansion history of the universe, aside from a relatively small preference for $w_{\rm DM}(z) > 0$ slightly above the 68\% CL around $z \sim 0.9$.

We then combined the two different versions of DESI BAO with CC+Pantheon+, considering the two distinct values of $r_{\rm d}$. The right column (excluding the topmost plot) of Fig.~\ref{fig:wdm_gpr-kernel} corresponds to these reconstructions of $w_{\rm DM}(z)$. We start with the results from CC+DESI-DR1+Pantheon+. For $r_{\rm d} = 149.3 \pm 2.7$ Mpc, we notice that in the high-redshift regime, there  is a mild preference of $w_{\rm DM}(z) < 0$ at slightly more than 68\% CL. On the other hand, for $r_{\rm d} = 137 \pm 3.6$ Mpc, CC+DESI-DR1+Pantheon+ predicts $w_{\rm DM}(z) < 0$ at approximately 68\% CL in the high-redshift regime, and $w_{\rm DM}(z) > 0$ at approximately 99\% CL in the intermediate-redshift regime ($z \sim 0.6$), but no evidence for $w_{\rm DM}(z) \neq 0$ at low redshifts is found in this case.
Replacing DESI-DR1 with DESI-DR2 does not lead to any significant changes; the results for the CC+DESI-DR1+Pantheon+ and CC+DESI-DR2+Pantheon+ combinations remain nearly identical.\footnote{We would like to point out that the deviation in $w_{\rm DM}(z)$, even though statistically mild, observed across all combined datasets including DESI BAO, occurs in a similar redshift range where DESI~\cite{DESI:2024mwx,DESI:2025zgx} reported a deviation from the $\Lambda$CDM model. This connection is noteworthy because, similar to the DESI analyses~\cite{DESI:2024mwx,DESI:2025zgx}, the present study adopts a model-independent diagnostic to investigate possible departures from $\Lambda$CDM cosmology, focusing on the DM sector.}

The overall results  indicate that the choice of the sound horizon scale $r_{\rm d}$ plays a significant role in the reconstruction process. As demonstrated by our analysis, the inferred behavior of cosmological parameters—particularly the DM EoS $w_{\rm DM}(z)$—varies with the assumed value of $r_{\rm d}$.

To systematically investigate this dependence, we evaluated $w_{\rm DM}(z = 0)$ over a range of $r_{\rm d}$ values within the interval $[135, 153]$~Mpc, using the combined datasets CC+DESI-DR1+Pantheon+ and CC+DESI-DR2+Pantheon+. Unlike the previous cases, where we accounted for the uncertainties in $r_{\rm d}$ associated with its two measurements from Refs.~\cite{eBOSS:2020yzd,Arendse:2019hev}, here we consider several discrete values of $r_{\rm d}$ in the interval $[135, 153]$~Mpc to obtain a rough estimate of $w_{\rm DM}(z = 0)$.
The results are summarized in Fig.~\ref{fig:whisker-GPR}, which shows the 68\% CL  constraints on $w_{\rm DM}(z = 0)$ in the form of a whisker plot. The figure illustrates that, across the range of $r_{\rm d}$, the constraints on $w_{\rm DM}(z = 0)$ consistently deviate from the CDM  assumption ($w_{\rm DM}(z) = 0$) if $r_{\rm d}$ starts to differ from $\sim 145$~Mpc, reinforcing the idea that $r_{\rm d}$ plays a non-negligible role in probing potential deviations from the standard CDM scenario. 

This sensitivity arises from the fact that $r_{\rm d}$ sets the physical scale of the BAO feature, which serves as a cosmological standard ruler. Since BAO measurements are used to calibrate distance scales and infer the expansion history of the universe, any assumption or uncertainty in $r_{\rm d}$ propagates directly into the reconstructed dynamics. Moreover, a non-zero $w_{\rm DM}(z)$ modifies the redshift evolution of the DM density, thereby influencing both the background expansion and the growth of large-scale structure. As a result, variations in $r_{\rm d}$ can either mimic or obscure the physical signatures of a non-zero $w_{\rm DM}(z)$. This interplay highlights the importance of carefully accounting for $r_{\rm d}$ when using BAO-based datasets to constrain the nature of DM and assess deviations from the CDM paradigm. 
In particular, as shown in Fig.~\ref{fig:whisker-GPR}, a mild departure of $w_{\rm DM}(z)$ from $w_{\rm DM} = 0$ is more pronounced for lower values of $r_{\rm d}$ compared to higher ones; however, departures in both directions are observed. 
A lower (higher) value of $r_{\rm d}$ implies a relatively higher (lower) value of $H_0$~\cite{Arendse:2019hev}. 
Since in this work we assume that the DE sector is represented by a cosmological constant and does not interact with the DM sector, an increased (reduced) value of $H_0$ corresponds to a modification of the standard $\Lambda$CDM model, specifically in its DM sector. 
Nevertheless, this mild departure should be interpreted with caution, as there exists a degeneracy between the late-time expansion history of the Universe and the value of $r_{\rm d}$, which is fixed by hand when incorporating the DESI BAO measurements. 
Therefore, variations in $r_{\rm d}$ are degenerate with the effects on the reconstructed $w_{\rm DM}$, and such variations are of the same order as the mild departures observed in $w_{\rm DM}(z=0)$.

\begin{table*}[t]
\caption{The table summarizes the mean values and the standard deviations.}
\footnotesize
\scalebox{1.0}{%
\begin{tabular}{ccccccc} 
\cline{1-7}\noalign{\smallskip}
\vspace{0.15cm}
Model & Datasets & $h$ &  $\Omega_{m}$  & $w_{\rm DM} (z=0)$
& $\ln B_{\Lambda \text{CDM},i}$  &  $-2\Delta\ln \mathcal{L_{\rm max}}$ \\
\hline
\hline
\vspace{0.15cm}
$\Lambda$CDM & CC &  0.676 (0.044) & 0.332 (0.063) & $0$ & $-$  &  $-$  \\
\vspace{0.15cm}
eosDM         & CC &  0.749 (0.078) & 0.229 (0.092) & $-$1.4 (2.4) & 3.41 (0.17)  & $-3.12$ \\
\hline
\vspace{0.15cm}
$\Lambda$CDM & Pantheon+ &  0.65 (0.14) & 0.331 (0.018) & $0$ & $-$  &  $-$  \\
\vspace{0.15cm}
eosDM         & Pantheon+ &  0.61 (0.14) & 0.25 (0.11) & 0.23 (0.51) & 6.11 (0.17)  &  $-2.62$ \\
\hline
\vspace{0.15cm}
$\Lambda$CDM & CC+Pantheon+ &  0.676 (0.028) & 0.329 (0.017) & $0$ & $-$  &  $-$  \\
\vspace{0.15cm}
eosDM         & CC+Pantheon+ &  0.663 (0.029) & 0.26 (0.11) & 0.31 (0.48) & 8.14 (0.19)  & $-0.45$ \\
\hline
\vspace{0.15cm}
$\Lambda$CDM & CC+DESI &  0.699 (0.025) & 0.286 (0.032) & $0$ & $-$  &  $-$  \\
\vspace{0.15cm}
eosDM         & CC+DESI &  0.685 (0.065) & 0.234 (0.088) & 0.16 (0.88) & 7.72 (0.39)  & $-3.04$ \\
\hline
\vspace{0.15cm}
$\Lambda$CDM & DESI+Pantheon+ &  0.719 (0.045) & 0.324 (0.018) & $0$ & $-$  &  $-$  \\
\vspace{0.15cm}
eosDM         & DESI+Pantheon+ &  0.63 (0.13) & 0.176 (0.066) & 0.85 (0.44) & 6.61 (0.39)  &  $-6.58$ \\
\hline
\vspace{0.15cm}
$\Lambda$CDM & CC+DESI+Pantheon+ &  0.689 (0.022) &  0.321 (0.016) & $0$ & $-$  &  $-$  \\
\vspace{0.15cm}
eosDM         & CC+DESI+Pantheon+ &  0.658 (0.028) & 0.229 (0.068) & 0.46 (0.35) & 9.03 (0.43) & $-3.53$ \\
\hline
\vspace{0.15cm}
$\Lambda$CDM & CC+DESI-DR2 &  0.671 (0.011) & 0.323 (0.027) & $0$ & $-$  &  $-$  \\
\vspace{0.15cm}
eosDM         & CC+DESI-DR2 &  0.711 (0.071) & 0.277 (0.063) & $-$0.9 (1.2) & 7.81 (0.25)  & $-1.15$ \\
\hline
\vspace{0.15cm}
$\Lambda$CDM & DESI-DR2+Pantheon+ &  0.702 (0.025) & 0.325 (0.018) & $0$ & $-$  &  $-$  \\
\vspace{0.15cm}
eosDM         & DESI-DR2+Pantheon+ &  0.60 (0.19) & 0.237 (0.098) & 0.18 (0.40) & 6.07 (0.23)  &  $-5.08$ \\
\hline
\vspace{0.15cm}
$\Lambda$CDM & CC+DESI-DR2+Pantheon+ & 0.693 (0.018) &  0.323 (0.017) & $0$ & $-$  &  $-$  \\
\vspace{0.15cm}
eosDM         & CC+DESI-DR2+Pantheon+ &  0.668 (0.029) & 0.302 (0.037) & 0.00 (0.21) & 9.55 (0.25) & $-2.67$ \\
\hline
\hline
\end{tabular}}
\label{tabla_evidencias}
\end{table*}
\begin{figure*}
   \includegraphics[trim = 0mm  0mm 0mm 0mm, clip, width=18.cm, height=18.cm]{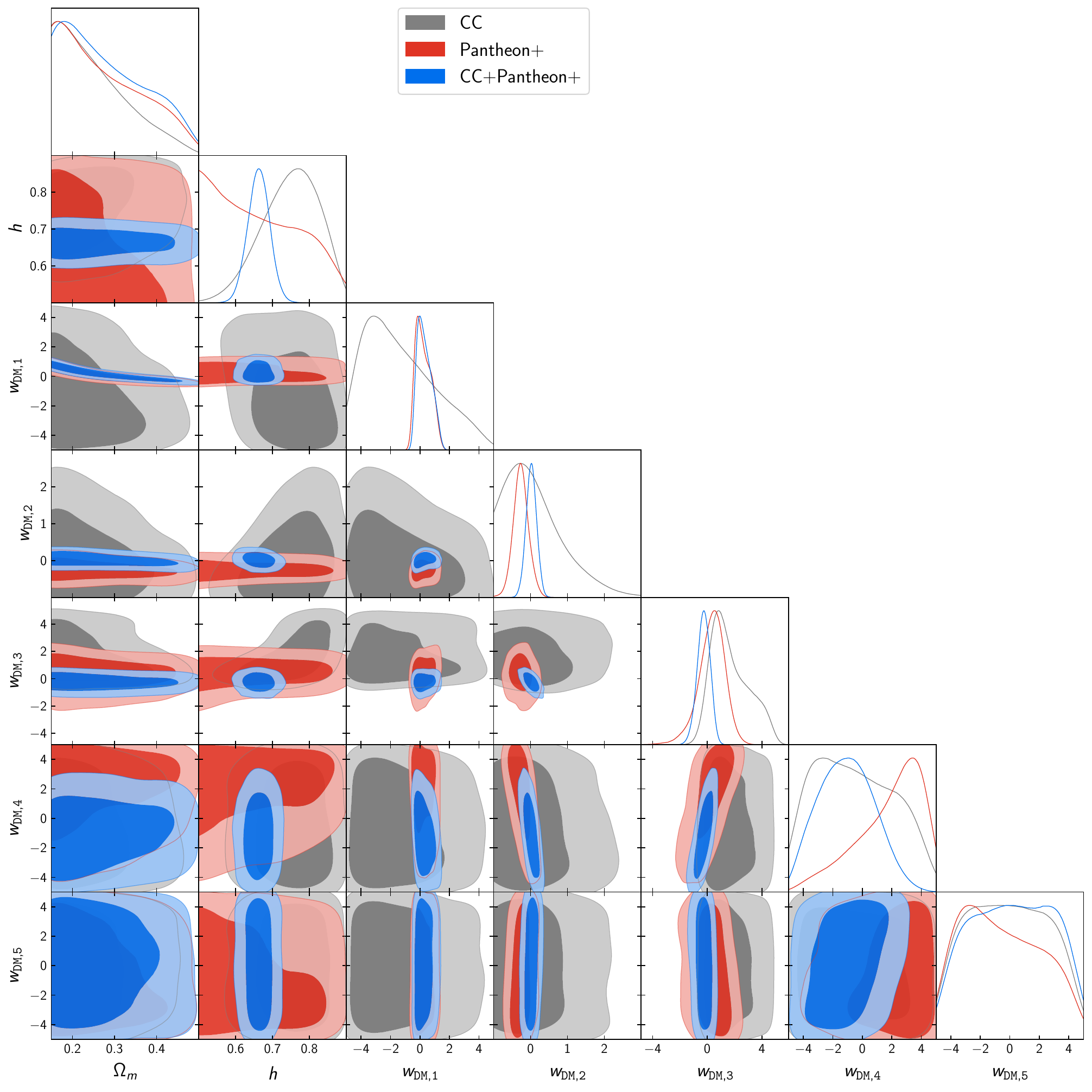}
   \caption{Triangle plot of the parametric reconstructions using three datasets without BAO. Shown are the marginalized 1D and 2D posteriors, illustrating their behavior within a specific cluster of dataset combinations.}  
   \label{fig:triangleplot}
\end{figure*}
\begin{figure*}
   \includegraphics[trim = 0mm  0mm 0mm 0mm, clip, width=18.cm, height=18.cm]{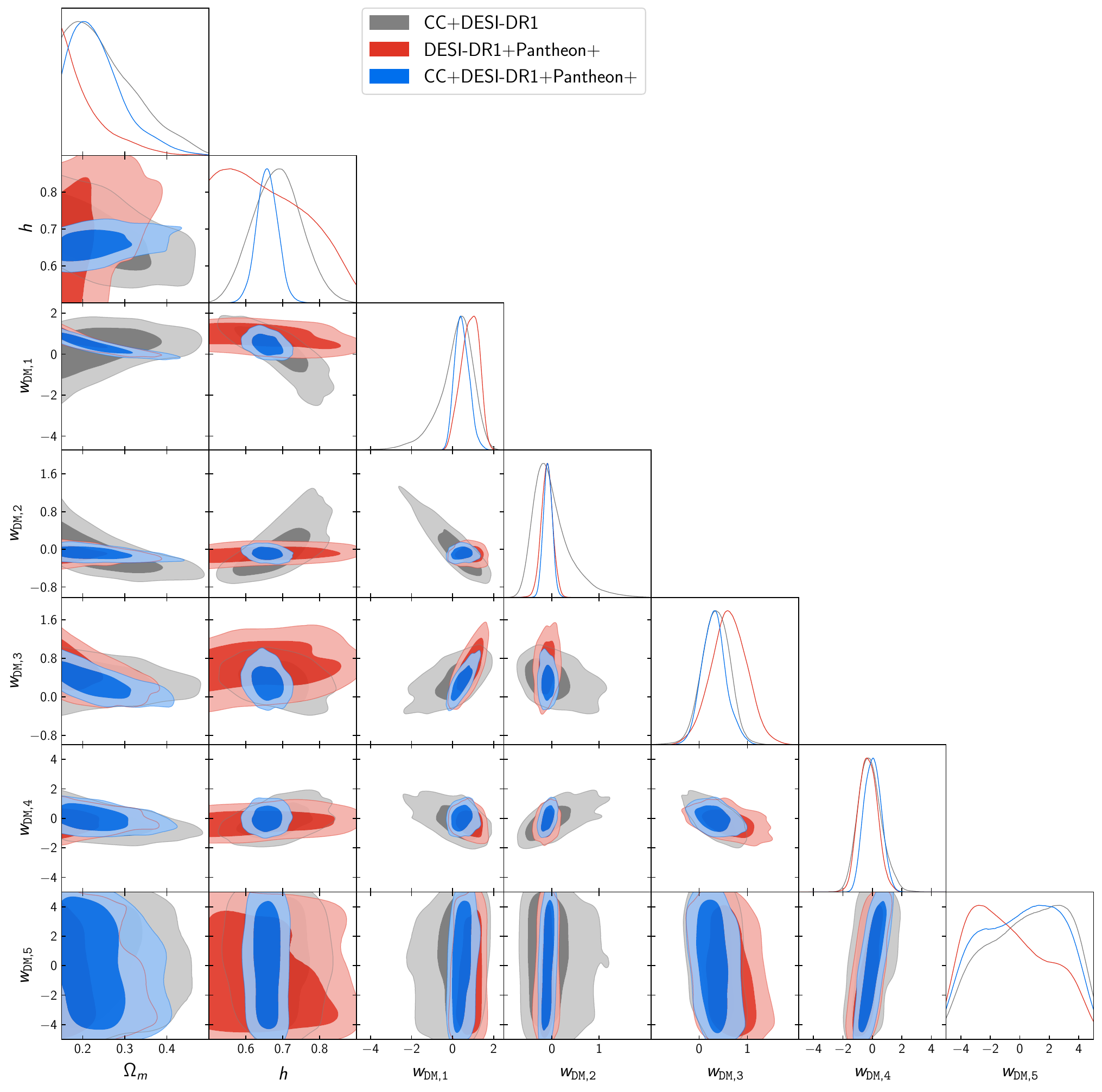}
  \caption{Triangle plot of the parametric reconstructions using three datasets including DESI-DR1 BAO. Shown are the marginalized 1D and 2D posteriors, illustrating their behavior within a specific cluster of dataset combinations.}
   \label{fig:triangleplot_desi}
\end{figure*}
\begin{figure*}
   \includegraphics[trim = 0mm  0mm 0mm 0mm, clip, width=18.cm, height=18.cm]{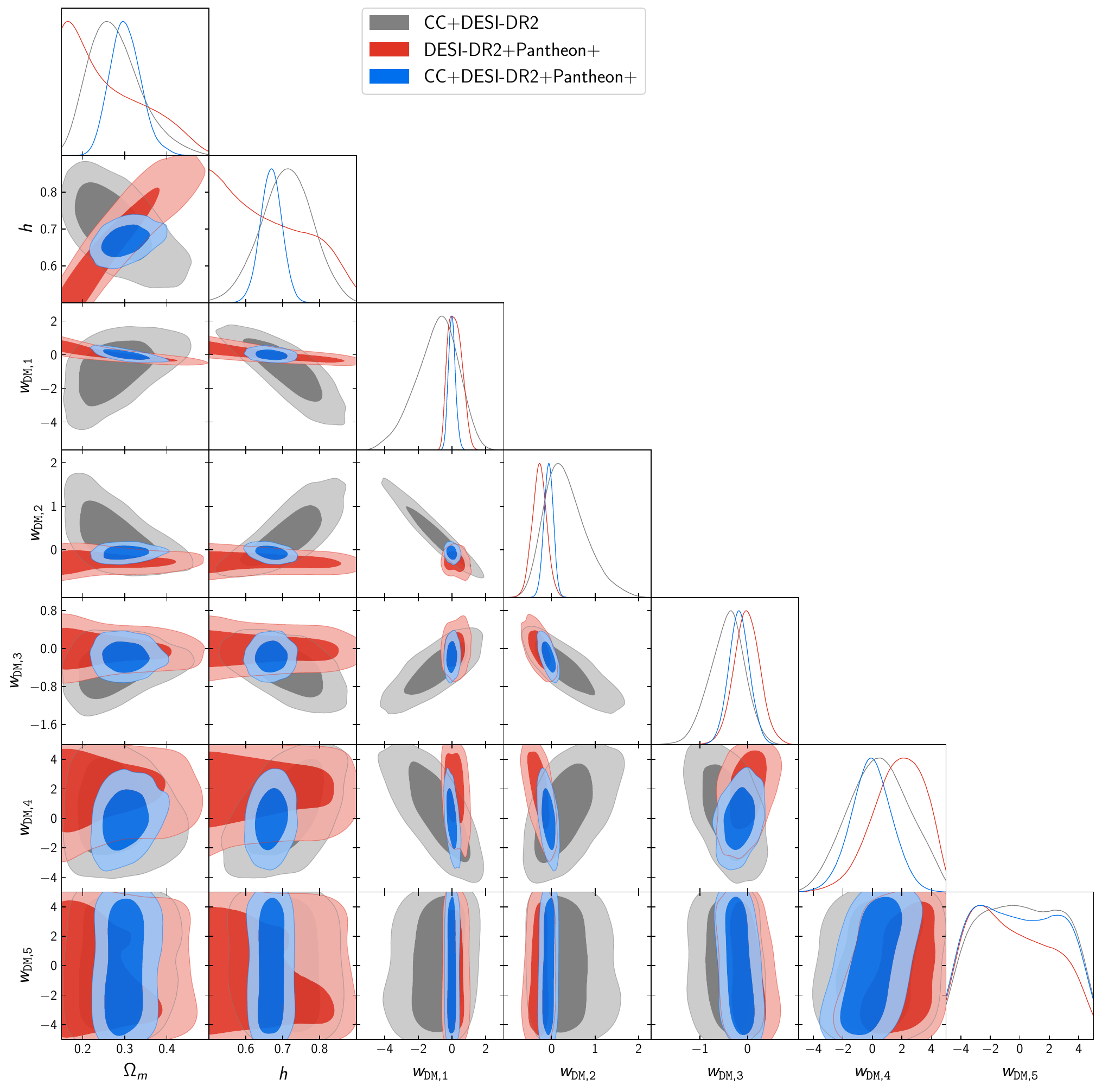}
   \caption{Triangle plot of the parametric reconstructions using three datasets including DESI-DR2 BAO. Shown are the marginalized 1D and 2D posteriors, illustrating their behavior within a specific cluster of dataset combinations.} 
   \label{fig:triangleplot_desi_dr2}
\end{figure*}

\subsubsection{Beyond the squared exponential kernel}

To assess the robustness of our results with respect to the choice of kernel, we also performed reconstructions using alternative covariance kernels beyond the standard squared exponential one, namely the Matérn family, Cauchy, and Rational Quadratic kernels.
For consistency, we used DESI DR1 and DR2, along with the two values of the sound horizon $r_{\rm d}$ mentioned in previous sections. However, the results for DESI DR1 and DESI DR2 are not significantly different; therefore, in this section we present only the results for DESI DR2.

To avoid overcrowding the manuscript with multiple figures, in Fig.~\ref{fig:deviation-various-kernels} we summarize the results for CC, Pantheon+, CC+Pantheon+, CC+DESI-DR2, DESI-DR2+Pantheon+, and CC+DESI-DR2+Pantheon+. 
In particular, in Fig.~\ref{fig:deviation-various-kernels} we present the quantity 
$\Delta w_{\rm DM}(z) = |w_{\rm DM}(z) - 0| / \sigma_{w_{\rm DM}(z)}$, 
which quantifies the statistical significance of the deviation of $w_{\rm DM}(z)$ from the CDM baseline ($w_{\rm DM} = 0$). It is important to emphasize that $\Delta w_{\rm DM}$ does not represent the absolute deviation, but rather the deviation normalized by its corresponding uncertainty, providing a clearer measure of statistical relevance across different kernels. According to the results shown in this figure, indications of a non-zero $w_{\rm DM}(z)$ are also supported beyond the {\it squared exponential} kernel. In the following, we present these results in detail.

We first focus on the reconstructions considering CC alone and Pantheon+ alone for all the kernels. For CC alone, we notice that in the low-redshift regime, $w_{\rm DM}(z)$ is almost identical to its null value, but in the high-redshift regime, a mild tendency for a non-null $w_{\rm DM}(z)$ within ($1\sigma$, $2\sigma$) is found. For Pantheon+ alone, irrespective of the kernel, we find that evidence for $w_{\rm DM}(z) \neq 0$ appears up to $2\sigma$, and, except for the {\it squared exponential} and Mat\'ern kernels, $w_{\rm DM}(z = 0)$ is found to be non-null at more than $1\sigma$.
Combining CC and Pantheon+ results in evidence of a non-null $w_{\rm DM}$ remaining within $2\sigma$ throughout the expansion history of the universe. In addition to that, only for the Mat\'ern $9/2$ kernel do we obtain $w_{\rm DM}(z = 0) > 0$ at slightly more than $1\sigma$.

When CC is combined with DESI-DR2, a preference for a non-zero $w_{\rm DM}(z)$ is found across all kernels, regardless of the chosen value of $r_{\rm d}$. Notably, this preference persists throughout the redshift range, although differences emerge at the present epoch, where various kernels yield differing levels of significance. For instance, the Mat\'ern family of kernels exhibits $w_{\rm DM}(z = 0) \neq 0$ at the $2$--$3\sigma$ level, with the Mat\'ern $9/2$ kernel in particular showing a deviation at nearly $3\sigma$. In contrast, the other kernels indicate $w_{\rm DM}(z = 0) \neq 0$ only at approximately the $1\sigma$ level.

For both DESI-DR2+Pantheon+ and CC+DESI-DR2+Pantheon+, we observe a hint of $w_{\rm DM} \neq 0$ throughout the expansion history of the universe. At the present epoch, all kernels yield a non-zero $w_{\rm DM}$ at approximately the $1\sigma$ level. More notably, in the intermediate redshift range, deviations reach up to $3\sigma$, suggesting evidence for a dynamical DM component. 
Additionally, we highlight that for the Mat\'ern $5/2$ and Mat\'ern $7/2$ kernels, the error estimates appear to be overestimated in the cases of the CC+Pantheon+, CC+DESI-DR2, and CC+DESI-DR2+Pantheon+ datasets. This behavior, which is not as pronounced in the other kernels, was also noted in~\cite{Seikel:2013fda}. 
Taken together, these results suggest that the possibility of a dynamical DM component deserves further investigation.

\subsection{Parametric approach}

We now turn our attention to the results of the parametric interpolation using GPR. Similar to the previous section, we consider nine different data combinations. The functional posteriors for the cases without and with DESI DR1 and DESI DR2 are presented in Fig.~\ref{fig:wdm_gp}.
In Table~\ref{tabla_evidencias}, we report the best-fit values of $h$ and $\Omega_{m,0}$, along with the natural logarithm of their Bayes factor and the quantity $-2\Delta\ln \mathcal{L}_{\rm max}$. The latter, if negative, indicates how well the fit to the data compares to the standard model. The 1D and 2D marginalized posteriors are shown in Figs.~\ref{fig:triangleplot}, \ref{fig:triangleplot_desi}, and \ref{fig:triangleplot_desi_dr2}. In the following, we examine each case in detail.

For the case where only the CC dataset is used to reconstruct $w_{\rm DM}$, we observe an expected improvement in $-2\Delta\ln \mathcal{L}_{\rm max}$ due to the additional parameters with respect to $\Lambda$CDM. The Bayes factor indicates moderate evidence against this reconstruction, which aligns with expectations. Notably, allowing $w_{\rm DM}$ to vary leads to a significant increase in the best-fit $h$ (see Table~\ref{tabla_evidencias}), though it remains largely unconstrained. This alleviates the Hubble tension but comes at the cost of reduced constraining power.
As shown in the upper part of Fig.~\ref{fig:wdm_gp}, the DM EoS exhibits an oscillatory behavior at the $1\sigma$ level, which is also reflected in the best-fit curve (red-dotted line).

For the Pantheon+ dataset, the larger data sample results in fewer oscillations compared to CC, though slight deviations from $\Lambda$CDM's $w_{\rm DM} = 0$ persist at the $1\sigma$ level. We also observe $w_{\rm DM} > 0$ around $z \approx 2.3$, followed by a sharp decline toward $z \approx 3$, mirroring the behavior seen with the kernel approach. We attribute this similarity and behaviour to the lack of data, and consequently of constraining power, in this redshift range, which causes the node located at $z = 3$ to default to a lower value than the previous one. 
While the parametric $w_{\rm DM}$ improves the fit, the Bayes factor suggests even weaker statistical support compared to the CC-only case. Moreover, both $H_0$ and $\Omega_m$ remain largely unconstrained, as indicated by the red contours in Fig.~\ref{fig:triangleplot}. In the standard model, the Pantheon+ dataset alone provides a relatively tight constraint on $\Omega_m$, yielding $\Omega_m = 0.332 \pm 0.018$. In contrast, the parametric scenario results in a significantly broader constraint, $\Omega_m = 0.249 \pm 0.107$.

This weakened constraint on $\Omega_m$ appears across all dataset combinations, although its magnitude varies depending on the case. 
It can be directly attributed to the additional degrees of freedom introduced by allowing the DM equation of state to vary. 
In the parametric approach, the dimensionless Hubble parameter is obtained after inference as  
\begin{equation}
    E^2(z) = \Omega_m(z) + \Omega_{\rm DE}(z),
\end{equation}
where we restrict our analysis to late-time redshifts (neglecting $\Omega_r \simeq 0$) and assume a spatially flat Universe ($\Omega_k = 0$).  
For $\Lambda$CDM, we have $w_{\rm DM} = 0$, and consequently $\Omega_m (z)  = \Omega_{m}(1+z)^3$ ($\Omega_m$ refers to the present day value of $\Omega_{m} (z)$). 
In contrast, for our parametric reconstruction, the situation is less straightforward, since $w_{\rm DM}$ is now a function of redshift. 
If only one bin (or node) were used, with a constant $w_{\rm DM,0}$, the matter density would evolve as  
\begin{equation}
\Omega_m (z) = \Omega_{m}(1+z)^{3(1+w_{\rm DM,0})},
\end{equation}
which introduces an additional degree of freedom, allowing $\Omega_m$
to vary in $[0.1, 0.5]$, since DM is no longer restricted to evolve strictly as $a^{-3}$. 
In our case, however, we use five bins (or nodes), which further increases this freedom. 
As a result, if the datasets employed lack sufficient constraining power, one should not expect $\Omega_m$ 
to behave as tightly as in the standard $\Lambda$CDM scenario.

Next, we analyze the combination of the CC and Pantheon+ datasets. This joint dataset considerably improves the constraints on the functional posterior, particularly within the range $0.5 < z < 1.5$ (upper-right panel of Fig.~\ref{fig:wdm_gp}). 
At higher redshifts, the mild preference for negative values becomes more evident. 
Nevertheless, despite the improvement relative to the individual CC and Pantheon+ cases, the confidence contours remain broad due to the scarcity of data in this regime. 
The constraining power is relatively low for CC, as shown in Fig.~\ref{fig:triangleplot}, where the dominant contribution to the constraints on the DM EoS parameters $w_{\rm DM,i}$ arises from Pantheon+. 
Interestingly, however, some influence from CC appears when considering the fourth node: since Pantheon+ contains only a handful of supernovae above $z = 1.5$, the few CC points available in this range can modestly affect the behavior of the node at $z = 2.25$. 
Although the CC dataset does not strongly impact the $w_{\rm DM,i}$ parameters, it provides something that Pantheon+ cannot: an anchor for $H_0$.

The best-fit curve (red-dotted line in the lower panel of Fig.~\ref{fig:wdm_gp}) trends toward negative values for $z > 1.2$, closely resembling the results obtained using the kernel approach with CC data alone. Notably, the improvement in fit, as measured by $-2\Delta\ln \mathcal{L}_{\rm max}$ (Table~\ref{tabla_evidencias}), is minimal compared to the standard model. In the well-constrained range $0.5 < z < 1.5$, $w_{\rm DM}$ gravitates toward zero, leading to negligible gains in data fitting. Given the additional parameters, this results in a significantly worse Bayes factor.

The next three cases will be discussed together, as they closely resemble the previously analyzed scenarios but now include DESI-DR1 BAO data. In the CC+DESI-DR1 case, the fit to the data remains comparable to its counterpart without DESI-DR1. However, this is not true for the other two cases. Both DESI-DR1+Pantheon+ and CC+DESI-DR1+Pantheon+ exhibit an overall improvement in data fitting. Despite this, the Bayes factor strongly favors the standard model, with an even greater preference than in cases without DESI-DR1.
Specifically, the Bayes factor values are approximately $7.7$, $6.6$, and $9$ for CC+DESI-DR1, DESI-DR1+Pantheon+, and CC+DESI-DR1+Pantheon+, respectively. Substituting DESI-DR1 with DESI-DR2 results in an overall poorer fit to the data (compared to $\Lambda$CDM) than in the DESI-DR1 cases; however, the Bayesian evidences remain largely unchanged.

Examining Fig.~\ref{fig:wdm_gp}, we observe a clear impact from incorporating DESI data. This inclusion leads to tighter constraints (i.e., smaller contours), while still preserving an oscillatory-like behavior across all six cases. Regarding these oscillations, we note that an interesting feature emerges in the best-fit reconstructions (red-dotted lines): a transition in behavior around $z \sim 1.5$. Specifically, when using DESI DR1, $w_{\rm DM}$ tends to favor positive values, whereas with DESI DR2, the preference shifts toward neutral or negative values. This behavior is readily explained by the differences between the two data releases, particularly the reduced error bar on the ELG2 point at $z \sim 1.32$ and the addition of a new QSO point at $z = 1.48$—precisely the redshift region where this transition occurs.

Another noteworthy characteristic is the best-fit value of $w_{\rm DM}$ at $z = 0$. Unlike the non-parametric reconstructions, which tend to favor negative values, these results generally indicate a preference for positive values, except in the case where only CC data is used. This is a direct consequence of treating $H_0$ (specifically, $h$) as a free parameter while fixing $r_{\rm d}$ (at $149.3$~Mpc) for the DESI BAO data in the parametric approach. This results in lower values of $h$ and $\Omega_{m,0}$ and higher values of $w_{\rm DM}(z = 0)$.
Another key factor contributing to this behavior can be observed in the marginalized posteriors (Figs.~\ref{fig:triangleplot}, \ref{fig:triangleplot_desi}, and \ref{fig:triangleplot_desi_dr2}). Since $w_{\rm DM}$ is allowed to vary freely, the usual correlation between $h$ and $\Omega_{m,0}$ disappears, leading to weaker constraints on both parameters at the background level given the datasets used.
Additionally, a direct comparison between Figs.~\ref{fig:triangleplot}, \ref{fig:triangleplot_desi}, and \ref{fig:triangleplot_desi_dr2} highlights the impact of the DESI $D_{\rm M}$ data on the posteriors (both DR1 and DR2), particularly for the DM EoS parameters. The exception is $w_{{\rm DM},5}$, which remains largely unaffected due to its location at $z = 3$, where data is sparse. Moreover, a tendency for positive values of $w_{\rm DM}(z = 0)$ becomes the standard trend when DESI data is included. 
The mild preference observed in DESI (both DR1 and DR2) for a non-zero DM equation of state, $w_{\rm DM} \neq 0$, may be understood in the broader context of its tendency to favor deviations from the $\Lambda$CDM paradigm. 
Recent analyses have already reported evidence for a preference toward dynamical dark energy within DESI data~\cite{DESI:2024mwx,DESI:2025zgx}. 
In our framework, however, DE is fixed to a non-dynamical form, and thus the flexibility that DESI data appear to demand cannot be accommodated through the DE sector. 
As a consequence, this preference for dynamics is effectively absorbed by the DM sector, manifesting as a slight deviation of $w_{\rm DM}$ from zero. 
It is important to emphasize, however, that this interpretation does not exclude the possibility of genuine physical modifications to the DM sector or to the processes of structure formation, which remain viable avenues for further exploration beyond the standard $\Lambda$CDM framework.

\section{Summary and Conclusions}
\label{sec-summary}

The dark sector of the universe, comprised of DM and DE, collectively constitutes nearly 96\% of the total energy budget of the universe. In an effort to probe its nature, numerous astronomical missions have been conducted over the years, yet a definitive answer remains elusive to cosmologists.
In this article, we have reconstructed $w_{\rm DM}$ using two methodologies based on Gaussian Process Regression (GPR): one being a non-parametric approach, and the other a parametric framework. The DE sector is represented by the cosmological constant in this work. More complex DE models could be considered—since the fundamental nature of DE remains unknown—but we opted for the simplest scenario, focusing exclusively on DM.

We began by performing non-parametric GPR reconstructions of $w_{\rm DM}$, exploring different kernels and three datasets: CC, Pantheon+, and DESI BAO, along with their combinations. The results extracted from the reconstructions of $w_{\rm DM}$ are presented in Figs.~\ref{fig:wdm_gpr-kernel}, \ref{fig:whisker-GPR}, and \ref{fig:deviation-various-kernels}, which jointly indicate a preference for a non-null $w_{\rm DM}$ throughout the expansion history of the universe. In fact, such evidence may increase up to $3\sigma$ during certain epochs (see Fig.~\ref{fig:deviation-various-kernels}), with the important caveat that this also depends on the choice of $r_{\rm d}$ and kernel. Additionally, in some cases, we observe potential sign changes in the DM EoS, with multiple transitions occurring in specific instances (see Fig.~\ref{fig:wdm_gpr-kernel}).
We also note a mild tendency for $w_{\rm DM}$ to be negative at the present time, as indicated by the recent datasets. This tendency becomes more pronounced when DESI BAO is combined with CC and Pantheon+ (see Fig.~\ref{fig:whisker-GPR}). This could be a potential indication of alleviating the $H_0$ tension (see \cite{DiValentino:2021izs} for details), since an increased value of $H_0$ may be associated with a lower value of $r_{\rm d}$, consistent with the presence of a negative DM EoS at the present epoch. These findings raise important questions about the consistency between datasets regarding the background expansion history, particularly the BAO measurements from DESI, as they appear to be driving the dynamical behavior of $w_{\rm DM}$.

In the parametric reconstruction using GPR as an interpolation method and employing the Nested Sampling algorithm, $w_{\rm DM}$ is obtained for the same datasets and their combinations (see Fig.~\ref{fig:wdm_gp}), this time without fixing the parameter $H_0$ ($h$). 
Compared to the non-parametric approach, we observe notable differences, such as a preference for $w_{\rm DM}(z = 0) > 0$, which is likely a consequence of the absence of correlation between $H_0$ and $\Omega_{m,0}$. This lack of correlation also results in lower inferred values for both $H_0$ and $\Omega_{m,0}$.
Similar to the non-parametric case, we observe a tendency for $w_{\rm DM}$ to oscillate around zero. However, in this scenario, the statistical significance of the deviations is much lower, never exceeding the $2\sigma$ level. Moreover, we find that substituting DESI DR1 with DESI DR2 leads to a noticeable change in the oscillatory behavior of the DM EoS. We attribute this change to the improved constraints provided by DR2 and the inclusion of an additional data point at $z = 1.48$—this redshift being the one where the change in behaviour occurs.

Based on the results summarized above, no strong conclusive evidence for $w_{\rm DM} \neq 0$ is found throughout the entire expansion history of the universe. However, the hints of a dynamical $w_{\rm DM}$, including the possibility of negative values, merit further investigation. This is particularly relevant for assessing whether a time-varying DM EoS is truly required, especially at late times, given the current observational data. Since this study approaches $w_{\rm DM}$ in a non-parametric framework, we hope it provides a valuable contribution to ongoing efforts in modern cosmological research.

Looking ahead, upcoming cosmological surveys such as \textit{Euclid} \cite{Euclid:2024yrr}, the \textit{Nancy Grace Roman Space Telescope}~\cite{Zellem2022NancyGR}, \textit{Legacy Survey of Space and Time} (LSST)~\cite{LSSTDarkEnergyScience:2018jkl},  {\it and the gravitational waves data}~\cite{Cozzumbo:2024vxw}
are expected to significantly improve constraints on the expansion history of the universe and the growth of structure. These missions will provide high-precision measurements of Type Ia supernovae, weak lensing, galaxy clustering, and Baryon Acoustic Oscillations over a wide range of redshifts and cosmic volumes. In particular, the increased data density and reduced observational uncertainties will allow for more robust non-parametric reconstructions of cosmological functions such as $H(z)$ and $D_{\rm M}(z)$, thereby enabling tighter constraints on the dark matter equation of state $w_{\rm DM}(z)$. This will be especially important for testing the presence of a dynamical or non-zero $w_{\rm DM}$, potentially distinguishing between cold and non-cold dark matter scenarios with greater confidence.

\section*{Acknowledgments}
The authors sincerely thank the referee for many insightful comments which significantly improved the article. MA acknowledges the Research Fellowship (File No: 08/0155(17246)/2023-EMR-I) from the Council of Scientific and Industrial Research (CSIR), Govt. of India. 
LAE acknowledges support from the Scientific and Technological Research Council of T{\"u}rkiye (T{\"U}B{\.I}TAK) under Grant No.\ 124N627. 
EDV acknowledges support from the Royal Society through a Royal Society Dorothy Hodgkin Research Fellowship. 
SP acknowledges the financial support from the Department of Science and Technology (DST), Govt. of India under the Scheme  ``Fund for Improvement of S\&T Infrastructure (FIST)'' (File No. SR/FST/MS-I/2019/41). 
WY has been  supported by the National Natural Science Foundation of China under Grants No. 12175096, and Liaoning Revitalization Talents Program under Grant no. XLYC1907098.  We acknowledge computational cluster resources at the ICT Department of Presidency University, Kolkata.  This article is based upon work from the COST Action CA21136 - ``Addressing observational tensions in cosmology with systematics and fundamental physics (CosmoVerse)'', supported by COST - ``European Cooperation in Science and Technology''.

\bibliography{references}

\appendix
\section{Gaussian process regression}
\label{sec-app-A}

In this section we describe the details of the Gaussian process regression and the remaining kernels that we used in the article.  The procedure for the Gaussian process is the following: 
\begin{enumerate}
    \item Firstly, we assume that we have $n$  data points. We model the function $f(z)$ as a realization of a Gaussian process~\cite{2012JCAP...06..036S,Mukherjee:2021ggf}, meaning that its values at any finite set of input locations follow a multivariate Gaussian distribution:
    \begin{equation}\label{f_Gp}
    f \sim \mathcal{GP}\left( M, K(Z,Z) \right),
    \end{equation}
    where $M = [M(z_1), M(z_2), \dots, M(z_n)]$ is the mean vector, and $K(Z, Z)$ is an $n \times n$ covariance matrix. Here, $Z = \{z_i \mid i = 1, 2, \dots, n\}$ represents the input set.

    \item The locations at which we reconstruct the function are denoted by $Z^* = \{z^*_i \mid i = 1, 2, \dots, n^*\}$. At these points, the function values are represented as $f^* = [f^*_1, f^*_2, \dots, f^*_{n^*}]$, where $f^*_i = f(z^*_i)$. These values are jointly distributed according to a Gaussian process:
    \begin{equation}\label{eqn_f}
    f^* \sim \mathcal{GP}\left( M^*, K(Z^*, Z^*) \right),
    \end{equation}
    where $M^* = [M(z^*_1), M(z^*_2), \dots, M(z^*_{n^*})]$ is the prior mean vector at $Z^*$, and $K(Z^*, Z^*)$ is the $n^* \times n^*$ covariance matrix evaluated at $Z^*$. The observational data are assumed to be noisy realizations of the underlying function:
    \begin{equation}\label{eqn_y}
    y_i = f(z_i) + \epsilon_i,
    \end{equation}
    where $\epsilon_i$ is Gaussian noise. Therefore, the data vector $y = [y_1, y_2, \dots, y_n]$ is distributed as
    \begin{equation}\label{eqn_y_gp}
    y \sim \mathcal{GP}\left( M, K(Z, Z) + C \right),
    \end{equation}
    where $C$ is the noise covariance matrix. If the observational errors are uncorrelated, $C$ is diagonal; otherwise, it is a full covariance matrix.

    \item The covariance function $K$ contains hyperparameters that control the shape and scale of the reconstructed function. To determine their optimal values, we maximize the log marginal likelihood \cite{2012JCAP...06..036S,Mukherjee:2021ggf}, which is derived by assuming a Gaussian prior $f \mid Z, \sigma_f, \ell \sim \mathcal{GP}(M, K(Z,Z))$ and a likelihood $y \mid f \sim \mathcal{GP}(f, C)$.  The resulting expression is
    \begin{eqnarray}\label{log-marginal-p}
    \ln \mathcal{L} &=& \ln p(y \mid Z, \sigma_f, \ell) \nonumber\\
    &=& -\frac{1}{2}(y - M)^T \left[ K(Z, Z) + C \right]^{-1} (y - M) \nonumber\\
    && - \frac{1}{2} \ln \left| K(Z, Z) + C \right| - \frac{n}{2} \ln 2\pi.
    \end{eqnarray}
    The hyperparameters $\sigma_f$ and $\ell$ are optimized by maximizing this likelihood. 

    \item By combining equations~(\ref{eqn_f}) and~(\ref{eqn_y_gp}), we obtain the joint distribution of the training outputs $\bm{y}$ and the reconstructed function values $f^*$~\cite{2012JCAP...06..036S,Mukherjee:2021ggf}:
    \begin{equation}\label{joint_y_f^*}
    \begin{bmatrix}
    \bm{y} \\
    f^*
    \end{bmatrix}
    \sim \mathcal{GP} \left(
    \begin{bmatrix}
    M \\
    M^*
    \end{bmatrix},
    \begin{bmatrix}
    K(Z, Z) + C & K(Z, Z^*) \\
    K(Z^*, Z) & K(Z^*, Z^*)
    \end{bmatrix}
    \right).
    \end{equation}
    Conditioning this joint distribution on the observed data $\bm{y}$ yields the posterior distribution for the reconstructed function $f^*$:
    \begin{equation}\label{post}
    f^* \mid Z^*, Z, \bm{y} \sim \mathcal{GP} \left(
    \overline{f^*}, \, \text{cov}(f^*)
    \right),
    \end{equation}
    where the mean and covariance of the posterior distribution are given by
    \begin{equation}\label{gp-mean}
    \overline{f^*} = M^* + K(Z^*, Z) \left[ K(Z, Z) + C \right]^{-1} (\bm{y} - M),
    \end{equation}
    and
    \begin{align}
    \text{cov}(f^*) &= K(Z^*, Z^*) - K(Z^*, Z) \nonumber \\
    &\quad \times \left[ K(Z, Z) + C \right]^{-1} K(Z, Z^*). \label{gp-cov}
    \end{align}
    Thus, we obtain the posterior mean vector $\overline{f^*}$ and covariance matrix $\text{cov}(f^*)$ for the reconstructed function.
\end{enumerate}

The reconstruction of the function's derivative is also Gaussian.  
In Gaussian Processes, the derivative of the function follows a multivariate normal distribution, just as the function prediction does. Consequently, the prediction of the derivative, denoted by $f^{'*}$, is given by $f^{'*} = [f^{'}(z_1^{*}), f^{'}(z_2^{*}), f^{'}(z_3^{*}), \dots, f^{'}(z_n^{*})]$.
The reconstructed derivative $f^{'*}$ follows a multivariate Gaussian distribution with mean $M^{'*}$ and covariance $K^{''}(Z^*, Z^*)$. The joint distribution involving the observations, the reconstructed function, and its derivative is given by:
\begin{align}\label{joint_1st_deri} 
\begin{bmatrix}  
y \\  
f^* \\  
f^{'*}  
\end{bmatrix} 
&\sim \mathcal{GP} \left(  
\begin{bmatrix}  
M \\  
M^* \\  
M^{'*}  
\end{bmatrix}, \right. \nonumber\\  
&\left.  
\begin{bmatrix}  
K(Z, Z) + C & K(Z, Z^*) & K^{'}(Z, Z^*) \\  
K(Z^*, Z) & K(Z^*, Z^*) & K^{'}(Z^*, Z^*) \\  
K^{'}(Z^*, Z) & K^{'}(Z^*, Z^*) & K^{''}(Z^*, Z^*)  
\end{bmatrix}  
\right).  
\end{align}
The prediction of the derivative function's mean and covariance is given below:
\begin{equation}\label{post_deri}
f^{'*} \mid Z^*, Z, y \sim \mathcal{GP} \left(
\overline{f^{'*}},\, \text{cov}(f^{'*})
\right),
\end{equation}
\begin{equation}\label{gp-deri-mean}
\overline{\bm{f}^{'*}} = M^{\prime *} + K^{'}(Z^*, Z)\left[ K(Z, Z) + C \right]^{-1} (\bm{y} - M),
\end{equation}
and
\begin{align}
\text{cov}(f^{'*}) &= K^{''}(Z^*, Z^*) - K^{'}(Z^*, Z) \nonumber \\
&\quad \times \left[ K(Z, Z) + C \right]^{-1} K^{'}(Z, Z^*), \label{gp-deri-cov}
\end{align}
where $\text{cov}\left( \frac{\partial f_i}{\partial z_i}, \frac{\partial f_j}{\partial z_j} \right) = K^{''}(z^{*}_i, z^{*}_j) = \frac{\partial^2 K(z^{*}_i, z^{*}_j)}{\partial z^{*}_i \partial z^{*}_j}$ are the elements of the $n^* \times n^*$ covariance matrix $K^{''}(Z^*, Z^*)$.

This case also includes optimized hyperparameter values to determine the mean and covariance of the derivative function. Furthermore, we use Eq.~\eqref{log-marginal-p} to obtain the optimal hyperparameter values. The second-order derivative case follows a similar methodology. After reconstructing the function and its derivative, one can reconstruct $w_{\rm DM}(f, f^{'*}, f^{''*})$, where $f$ represents either $E$ or $D$, and its uncertainty can be estimated using the error propagation rule. For more details we refer to ~\cite{2012JCAP...06..036S}.

\section{Bayesian Statistics}
\label{sec-app-B}

To perform a parameter inference procedure, one applies Bayes' theorem:
\begin{equation}
    P(u|D,M) = \frac{\mathcal{L}(D|u,M)P(u|M)}{E(D|M)},
\end{equation}
where $u$ is the vector of parameters of model $M$, $D$ is the data, $P(u|D,M)$ is the posterior probability distribution, $\mathcal{L}(D|u,M)$ is the likelihood, $P(u|M)$ is the prior distribution, and $E(D|M)$ is the Bayesian evidence.
Once the Bayesian evidence is computed for two models, $M_1$ and $M_2$, the Bayes factor is defined as
\begin{equation}
    B_{12} \equiv \frac{E(D|M_1)}{E(D|M_2)}.
\end{equation}
By taking the natural logarithm of the Bayes factor and using the empirical revised Jeffreys' scale in Table~\ref{jeffreys}, we can quantify how much better (or worse) model $M_1$ is compared to model $M_2$ in explaining the data.

\begin{table}
\footnotesize
\scalebox{1.2}{%
\begin{tabular}{cc} 
\cline{1-2}\noalign{\smallskip}
\vspace{0.15cm}

$\ln{B_{12}}$ & Strength of evidence \\
\hline
\hline
\vspace{0.15cm}
$< 1.0$ & Inconclusive \\
\vspace{0.15cm}
$[1.0, 3.0]$ & Moderate evidence \\
\vspace{0.15cm}
$[3.0, 5.0]$ & Strong evidence \\
\vspace{0.15cm}
$> 5.0$ & Decisive evidence \\
\hline
\hline
\end{tabular}}
\caption{Revised Jeffreys' scale for model selection based on the logarithm of the Bayes factor $\ln B_{12}$, following the convention of~\cite{Kass:1995loi}.}
\label{jeffreys}
\end{table}

\end{document}